\documentclass[twoside]{bourlarge}
\usepackage{carpentier_h}

\usepackage{dsfont}

\usepackage{amssymb}
\usepackage{amsmath}
\usepackage{graphicx}

\newcommand\Bnabla[1]{{}^{#1 \hspace{-0.03cm}} \nabla }

\begin{document}
\Large

\title{Topology of Bands in Solids :\\
From Insulators to Dirac Matter}

\author{David {\sc Carpentier} \\
Laboratoire de Physique, \\
Ecole Normale Sup{\'e}rieure de Lyon \\ 
46, All{\'e}e d'Italie\\ 
69007 Lyon, France }

\maketitle

\begin{abstract}
Bloch theory describes the electronic states in crystals whose energies are distributed as bands over the Brillouin zone. The electronic states 
corresponding to a (few) isolated energy band(s) thus constitute a vector bundle. The topological properties of these vector bundles provide new characteristics of the 
corresponding electronic phases. We review some of these properties in the case of (topological) insulators and semi-metals. 
\end{abstract}



\section{Introduction}

Topology is a branch of mathematics aiming at  identifying properties of various objects invariant under continuous deformations. Examples range from the 
case originally considered by Euler of 
paths in the city of K\"onigsberg crossing all of its seven bridges, to the two classes of two dimensional surfaces characterized by an Euler topological index, 
and the problem of the hairy ball or M{\"o}bius strip related to the more abstract vector bundles considered in this paper. 
Topological considerations  occurred in different domains of physics, from the seminal work of P. Dirac on the quantization of magnetic monopoles 
  \cite{Dirac1931} to the classification of defects such as vortices, dislocations or skyrmions in ordered media  \cite{Toulouse:1976}. 
In the last ten years topology has arisen as a useful tool in the old domain of the quantum band theory of crystals. 
In this theory, single electron states are  determined from the geometry of the  lattice and the nature of atomic orbitals constituting the solid. 
These Bloch states are labelled by their quasi-momentum which belongs to the first Brillouin zone, a $d$-dimensional torus. 
Mathematically, the object defined out the ensemble of electronic states labelled by the momentum $k$ is a vector bundle over the Brillouin zone. Such objects are 
known to display possible non trivial topology, such as the above mentioned twisted M{\"o}bius strip or the hairy ball. 
In the context of band theory, this new characterization of an 
ensemble of eigenstates over the Brillouin torus has led to the notion of topological insulator. 
The full ensemble of states over the Brillouin torus is always trivial and lacks any topological property. However, it can be split into two well-separated sub-ensemble of states {\it e.g.} by an energy gap, both of which possessing a non trivial topology. This is the case of the valence and conduction bands of a topological insulator. 

 This notion of topological property of bands was first identified in the context of the two dimensional Integer Quantum Hall Effect soon after its discovery, 
 in the pioneering work of D. Thouless {\it et al.} \cite{Thouless:1982}. 
The initial framework of the band theory of crystals was found to be inadequate and restrictive (in particular translations on the crystal do not 
commute due to the magnetic field and require the definition of a magnetic unit cell) and soon the initial topological characterization evolved
 to a more general  form \cite{Avron:1985,Niu:1985}. In this context it later evolved as a property of various interacting electronic and magnetic 
 phases leading to the notion of topological order  \cite{Wen:1990}. 
  The initial requirement of a magnetic field was later lifted by D. Haldane. By considering a model of graphene with a time reversal symmetry breaking potential, 
  but a vanishing net flux through each unit cell, he preserved the translational symmetry of the original lattice \cite{Haldane88}. The original Chern topological 
  index of D. Thouless {\it et al.} \cite{Thouless:1982} was then understood as a topological property of bands in two dimensional insulators without 
 time reversal symmetry. These phases are now called Chern insulators. 
 From the beginning, this topological property was discussed in relation with geometrical phases acquired during
 an adiabatic evolution studied by M. Berry \cite{Berry1984}. Both were discussed as the parallel transport of eigenstates  following 
 similar ``Berry connexions'' \cite{Simon1983}. The topological Chern index describes the impossibility to perform such a parallel transport of Bloch states 
 globally over the Brillouin zone.

 The seminal work of  C. Kane and G. Mele completely renewed the interest on the topological characterization of bands in crystals and opened new 
 fascinating perspectives of experimental importance \cite{KaneMele2005}. The authors considered a model of graphene in the presence of a strong spin-orbit 
 interaction, which preserves time-reversal symmetry and discovered a insulating phase characterized by a new topological property, now called 
 the Quantum Spin Hall Effect. While their model takes its roots in the work by D. Haldane, the new topological index is not related in general to the previous 
 Chern index : it characterizes different topological properties. Indeed, the topological twist of a band probed by this Kane-Mele index is a consequence of the 
 constraints imposed on the electronic states by the time-reversal symmetry. It is a symmetry-related topological property. 
 The interest of this new band property grew considerably when it was realized that it existed also in bulk three dimensional materials. A simple recipe of 
 band inversion due to a strong spin orbit led to the proposal and discovery of this topological property in a large class of materials. 
How is this topological property detected ? 
 The hallmark of topological bands in an insulator is the existence of metallic  states at its surface which can be probed by various surface techniques. 
 Moreover, the surface states are typically described as Dirac particles, whether in one of two dimensions. Indeed, their  dispersion relation corresponds 
 to the linear crossing of two bands, which is described phenomenologically at low energy similarly to graphene 
 by a relativistic Dirac equation. The existence of these Dirac surface states implies  transport properties different from those of conventional metals, 
 stimulating a great number of theoretical and experimental work.

 The present review does not aim to cover the various aspects of topological insulating phases, nor to discuss exhaustively any of them. There already exist 
 several detailed reviews  \cite{HasanKane2010, Qi:2011} and textbooks \cite{Bernevig,FranzMolenkamp}. Instead we will give a synthetic overview of the salient 
 features of the topological properties of electronic bands. We will start by reviewing the basics of Bloch theory and the Berry phase as a notion of parallel transport 
 of electronic states. We will then discuss the specificities of this Berry formalism within  Bloch theory. The definition of the 
 Chern and Kane-Mele topological indices will be derived on simple historical graphene models. We will conclude by a discussion of Dirac fermions both as 
 surface states but also critical models from topological insulators and describe their topological properties.

\section{Bloch Theory}
\label{sec:Bloch}

\begin{figure}[htbp!]
\begin{center}
\includegraphics[width=5cm]{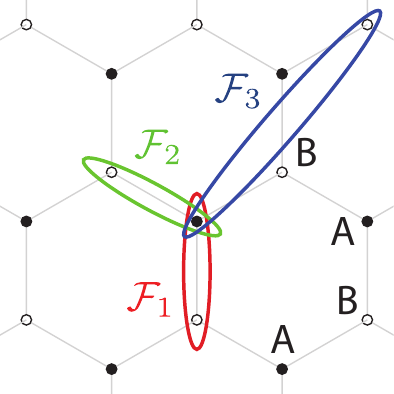}
\hspace{.3cm}
\includegraphics[width=5cm]{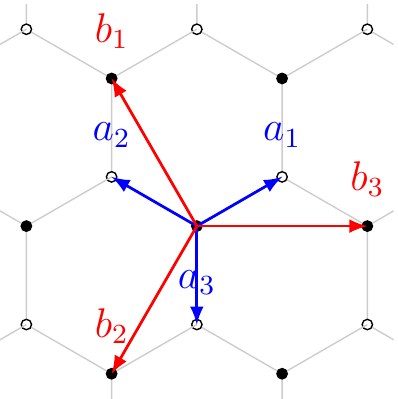}
\end{center}
\caption{Representation of the honeycomb lattice of Carbon atoms in graphene. This lattice is not a Bravais lattice, and possesses $2$ sub lattices $A$ and $B$, 
represented by black and open circles. 
Different inequivalent choices of fundamental domains $\mathcal{F}_i$ are shown. 
Vectors of the triangular Bravais  lattice are defined in red on the right figure. }
\label{fig:HoneycombLattice}
\end{figure}

 Crystalline solids are grossly classified into insulators and metals depending on their electronic transport properties. Within the band theory of crystals, this behavior depends on the existence of a gap between energy bands corresponding to occupied electronic states, and empty states for energies above the gap. 
   This notion of energy bands originates from Bloch theory  \cite{Blount1962}. Indeed, the description of the quantum states of electrons  in solids starts by 
  the identification  of conserved quantities and associated quantum numbers, in a standard fashion in  quantum mechanics. 
 In crystals, the  symmetry of the lattice implies that a discrete ensemble of translations commutes with the Hamiltonian, allowing to define a conserved pseudo-momentum $k$. 
 More precisely, we shall consider electronic wave functions restricted to a crystal $\mathcal C$, the locations of atoms, which constitutes a discrete ensemble of 
points in the Euclidean space  of dimension $d$.
  By definition, this crystal is invariant by an ensemble of discrete translations $T_\gamma$ of vectors $\gamma$ belonging to the so-called
 Bravais lattice $\Gamma$. 
 In general, the initial crystal may not be itself a Bravais lattice : 
 translations by only a subset of vectors connecting a pair of atoms does  leave the crystal invariant. 
 In that case, we can identify 
 $N>1$ sub-lattices defined as ensembles of points of the crystal $\mathcal C$ related by translations of the Bravais lattice. 
 All points of the crystal can then be deduced  by translations from points of a fundamental domain $\mathcal F$, 
also called a unit cell of the crystal.
Necessarily, the choice of a fundamental domain $\mathcal F$ amounts to choose one point in each sub-lattice. 
Of course any translation $\mathcal F+\gamma$ of a fundamental domain is also a possible choice for a fundamental domain. However, there exists (for $N>1$) 
different possible choices of $\mathcal F$ that are not related by translation : we will come back to this point later in our discussion. 
A canonical example in two dimensions of such a non-Bravais lattice is the honeycomb lattice of  $p_z$  orbitals of Carbon atoms in graphene which we will use 
throughout this article.  
It possesses $N=2$ sub-lattices, represented in Figure ~\ref{fig:HoneycombLattice}. 
Three different choices for fundamental domains for graphene are illustrated in  Fig.\ \ref{fig:HoneycombLattice}.

 Having determined the Bravais lattice $\Gamma$, we can now simultaneously diagonalize the hamiltonian $H$ defined on the crystal and the set of translation operators 
 $T_\gamma$ for $\gamma \in \Gamma$. 
 Eigenfunctions of the unitary operators $T_\gamma$ are Bloch functions, defined by their pseudo-periodicity 
 \begin{equation} 
 T_\gamma\psi(x)=\psi(x-\gamma) = e^{i k.\gamma} \psi(x) \textrm{ for any point } x\in \mathcal C,  
 \end{equation}
for some vector  $k \in \mathds{R}^{d}$. 
This property defines Bloch functions on $\mathcal C$ with 
quasi-momentum $k$.  The ensemble of Bloch functions with a fixed quasi-momentum form a finite-dimensional space 
$\mathcal{H}_k$ with the scalar product 
$\langle  \psi_k |  \chi_k\rangle_k=\sum_{x\in \mathcal C/\Gamma}\overline{\psi}_k(x) \chi_k(x)$. 
Note that since a Bloch function is uniquely determined by its value on 
a fundamental domain $\mathcal F$, we obtain $\dim(\mathcal{H}_k)=N\times \dim(V)$, where $N$ is the number of sub lattices.  
 At this stage, it is useful to introduce the reciprocal lattice $\Gamma^\star$   of the Bravais lattice $ \Gamma$, and 
 composed of vectors $G$ such that $G\cdot\gamma\in2\pi{\mathds Z}$ for any $\gamma \in \Gamma$. 
 As $\ee^{\ii k \cdot\gamma}=\ee^{\ii(k+G)\cdot\gamma}$ for $G\in \Gamma^\star$, the spaces $\mathcal{H}_k$ and $\mathcal{H}_{k+G}$ can be identified : 
 the quasi-momentum $k$ takes values only in the $d$-dimensional Brillouin torus $\BZ=\mathds{R}^{d}/\Gamma^\star$ obtained by identifying $k$ and $k+G$ for  
$G\in \Gamma^\star$. 

 The electronic wave functions we consider belong to the Hilbert space $\bf{H}=\ell^{2} (\mathcal C,V)$  of  functions defined on the crystal 
 $\mathcal C$, taking values in a finite dimensional complex vector space $V$ which accounts for the various orbitals per atom kept 
 in the description, and square-summable with  the scalar product 
 $\langle \psi|\chi\rangle =\sum_{x\in\mathcal C}\langle\psi(x)|\chi(x)\rangle_V$.  
  The identification of the initial Hilbert space $\bf{H}$ with the ensemble of Bloch functions $\mathcal{H}_k$ is provided by the standard Fourier transform. 
 From any such function $\psi\in \bf{H}$, the  Fourier transform 
\begin{equation}
\widehat\psi_k(x)=\sum_{\gamma\in\Gamma}\ee^{-\ii k\cdot\gamma}
\psi(x-\gamma), 
\label{FT}
\end{equation}
defines a function $\widehat\psi_k$ which is pseudo-periodic in the sense defined above : 
\begin{equation}
\widehat{T_\gamma\psi}_k(x)= \widehat\psi_k(x-\gamma)=\ee^{\ii k\cdot\gamma}\widehat\psi_k(x)  . 
\label{BF}
\end{equation}
Note that we also have $\widehat\psi_k=\widehat\psi_{k+G}$ for $G\in \Gamma^\star$. 
Hence the function $\hat{\psi}_k$ belongs to the space $\mathcal{H}_k$ of Bloch functions with pseudo-momentum $k$. 
The inverse Fourier transform is naturally given by the integration over the Brillouin torus:
\begin{equation}
\psi(x)=|\BZ|^{-1}\int_\BZ\widehat\psi_k(x)\,dk, 
\end{equation}
where $|\BZ|$ stands for the volume of the Brillouin Zone. 
This Fourier transform (\ref{FT}) is an isomorphism between the Hilbert space $\bf{H}$ and 
the ensemble $\{\mathcal{H}_k\}_{k\in \textrm{BZ}}$ of Bloch functions with momentum $k$ in the Brillouin torus, which allows to "identify" both spaces. 
The correspondence of norms is provided by the Plancherel formula. 
 The commutation of the Hamiltonian with the translations $T_\gamma$ for $\gamma \in \Gamma$ implies that the eigenstates $\psi^{\alpha}_k$ 
 are Bloch functions with $k$ a conserved quantity (quantum number). The evolution of an eigenenergy $\epsilon^\alpha_k$ with the quasi-momentum $k$ 
 defines an \emph{energy band}.  The number $M$ of energy bands is thus given by $\dim(\mathcal{H}_k)$ : the number of orbitals per lattice site times the 
 number $N$ of sub lattices. The nature of the electronic phase is then determined from both the band structure and the filling factor : 
 how many electrons are initially present per lattice site. Three different situations occurs : 
 \begin{description}
\item[Insulator :] an energy gap occurs between occupied states below the chemical potential and empty states above the chemical potentiel. This corresponds to an insulator, with a minimum energy necessary to excite electrons in the crystal
 \item[Metal :] the chemical potential crosses some energy bands over the Brillouin torus : no minimum exists for the excitation energy
 \item[Semi-Metal or Zero Gap Insulator :] two or more energy bands crosses exactly at (or close to) the chemical potential.  This is the situation of graphene. Close to this crossing, electronic excitation possess a linear dispersion relation, which mimics that of ultra-relativistic particles. 
  \end{description}
 In the following, we will be interested mostly by the behavior of the eigenstates~$\psi^{\alpha}_k$ for the occupied bands of a crystal, or the bands below the crossing point of semi-metals.  


\section{Geometrical Phase and Parallel Transport}

We now turn ourselves to a discussion of transport of Bloch functions as their momentum $k$ evolves in the Brillouin zone, starting from the initial notion 
of geometrical phase in quantum mechanics.

\subsection{Aharonov-Bohm Effect}

\begin{figure}[htbp!]
\begin{center}
\includegraphics[width=8cm]{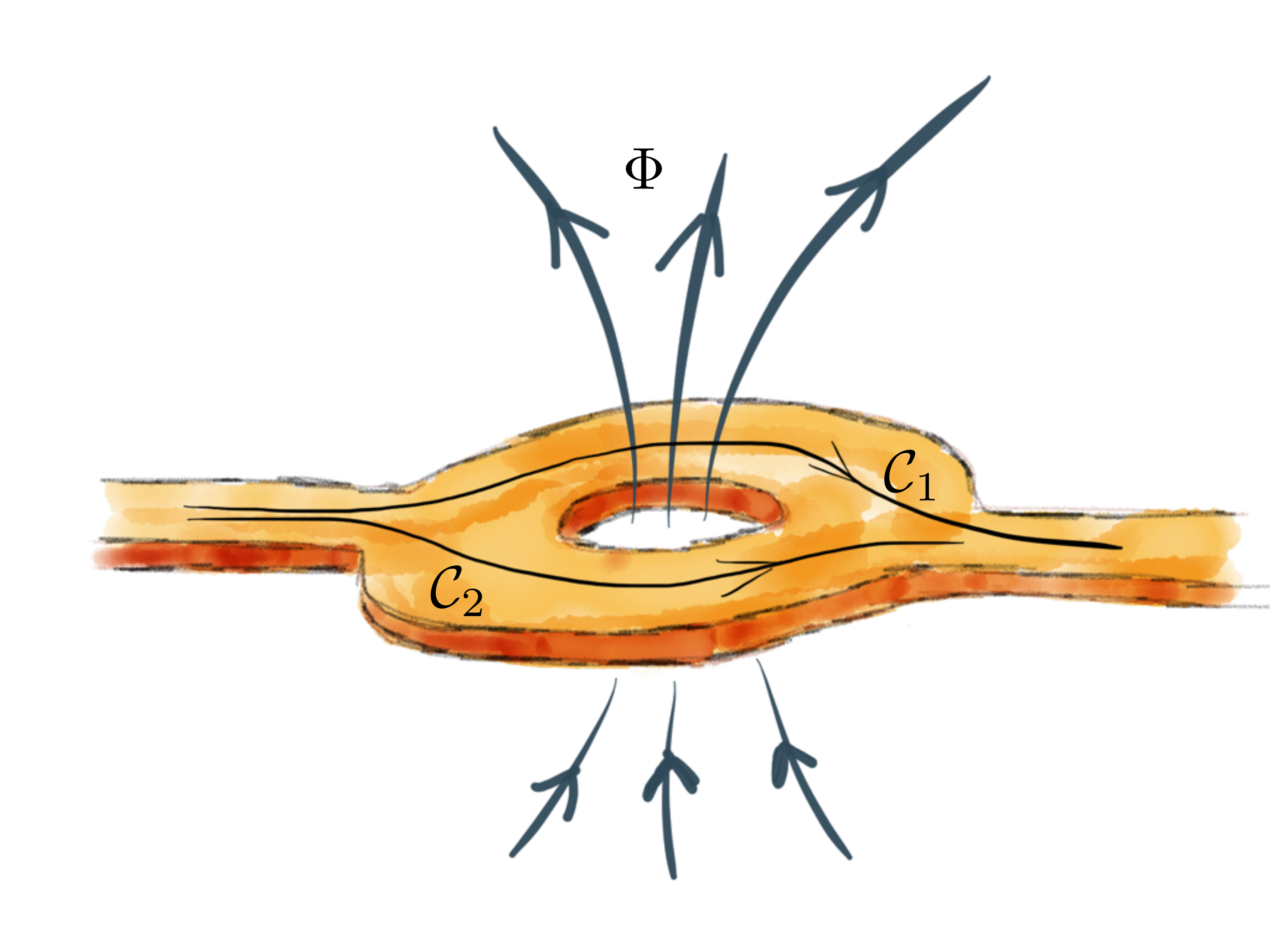}
\end{center}
\caption{Sketch of the geometry of  small metallic ring threaded by a magnetic flux, allowing to probe the Aharonov-Bohm effect. }
\label{fig:Aharonov-Bohm}
\end{figure}

Consider a small metallic ring threaded by a magnetic field localized  inside its hole (see Fig.~\ref{fig:Aharonov-Bohm}). 
We assume that the propagation of electrons can be considered to be phase coherent through the system. 
Even though the electrons propagating inside the conductor never encounter the magnetic field, they feel its presence through a dephasing effect : 
an electron propagating along a trajectory $\mathcal{C}$ acquires an extra phase induced by the electromagnetic potential $A(x)$ : 
\begin{equation}
\theta_{\mathcal{C}} = - \frac{e}{\hbar} \int_\mathcal{C} A(x).dx . 
\label{eq:AB}
\end{equation}
This is the so-called Aharonov Bohm effect \cite{Aharonov:1959}: 
 Two trajectories $\mathcal{C}_1$ and $\mathcal{C}_2$ going 
 from $x_1$ to $x_2$ will correspond to a respective dephasing 
 \begin{equation}
\theta_1 - \theta_2 =   \frac{e}{\hbar} \oint_{\mathcal{C}_1 U \mathcal{C}_2} A(x).dl
=  2\pi \frac{\phi}{\phi_0},  
\end{equation}
 where $\phi_0 = h/e$ is the flux quantum for electrons and $\phi$ is the magnetic flux enclosed between $\mathcal{C}_1$ and $\mathcal{C}_2$. 
  The modulation with the flux $\phi$ of the respective phase can be experimentally tested by monitoring the electronic current through such a small gold annulus
  as a function of the magnetic field threading the sample : oscillations  of periodicity $\phi_0$  are the hallmark of this Aharonov-Bohm effect \cite{Webb:1985}. 
 Note that although the physical quantity (the current) depends only on the gauge invariant magnetic flux, the evolution along a given path $\mathcal{C}$ depends 
 on the choice of gauge to express the vector potential $A(x)$. A consistent choice is required to be able to describe the evolution of electronic states 
 along any path.

\subsection{Berry phase}

Let us now discuss the notion of Berry phase. For the sake of pedagogy, we will consider the adiabatic evolution of eigenstates of a Hamiltonian
 following the initial work of M. Berry  \cite{Berry1984}  (see also \cite{Kato:1950}), although this notion extends to more general cyclic evolution of states 
 in Hilbert space  \cite{Aharonov:1987,Anandan:1990}. 
We consider a Hamiltonian $H(\lambda )$ parametrized by an ensemble of external parameters $\lambda=(\lambda_1,\lambda_2,...)$.  
These external parameters evolve in times.  This continuous evolution can be represented by a curve $\lambda(t), t_i\leq t \leq t_f$ 
in the parameter space. 
We study the time evolution of a given eigenstate $|\psi(\lambda)\rangle$ of $H(\lambda)$ with energy $E(\lambda)$. 
It is further assumed that this eigenstate is non-degenerate and that during the evolution, the energy $E(\lambda)$ is separated from all other energies,
 and furthermore that the system will remain in this eigenstate. 
With these assumptions, the evolution of the initial state takes place in the $1$-dimensional vector space of instantaneous eigenvectors of $H(\lambda )$ with 
energy $E(\lambda)$. For normalized states, the only remaining degree of freedom characterizing this evolution lies in the phase of the eigenstate. 
 
 To describe the evolution of this phase along the curve $\lambda(t)$ in parameter space, let us consider a ``continuous" basis of eigenstates 
 $\{ |\phi_n(\lambda)\rangle\}_n$  of the  Hilbert space. This basis is assumed continuous in the sense that the $\phi_n(\lambda ;x)$ are continuous wave functions 
 as $\lambda$ varies
 for any position $x$. By definition, the eigenstate $|\psi (\lambda(t))\rangle$ can be decomposed on this basis of eigenstates leading to  
 \begin{equation}
|\psi (\lambda(t)) \rangle = e^{-i\theta(t)} | \phi_{n_0}(\lambda(t)) \rangle, 
 \end{equation}
 which satisfies the instantaneous Schr\"odinger equation 
 \begin{align}
& H(\lambda(t))  \ \psi(\lambda(t)) \rangle   =  i \hbar ~\partial_t  | \psi(\lambda(t))  \rangle\\
\Rightarrow 
& \hbar ~ \partial_t \theta(t)= E_{n_0}(\lambda(t)) - i  ~   \langle \phi_{n_0}(\lambda(t)) | \partial_t   \phi_{n_0}(\lambda(t)) \rangle . 
 \end{align}
 The evolution of the phase $\theta(t)$ of the eigenstate is thus given by the sum of two terms : the usual dynamical phase $\theta_\textrm{dyn}$ 
 originating from the evolution of the energy of the state and a new contribution, denoted the geometrical Berry  phase $\theta_{n_0} $ : 
\begin{align}
& \theta(t_f) - \theta(t_i)  = \theta_\textrm{dyn} + \theta_\textrm{B} \\
& \theta_\textrm{dyn}   = \frac{1}{\hbar} \int_{t_i}^{t_f}  E_{n_0}(\lambda(t)) dt  
\quad ; \quad
\theta_{n_0} = 
- i \int_{t_i}^{t_f}     \langle \phi_{n_0}(\lambda(t)) | \partial_t   \phi_{n_0}(\lambda(t)) \rangle ~dt . 
\nonumber
 \end{align}
 This new contribution can be rewritten as a purely geometrical expression over the parameter space : 
 \begin{equation}
\theta_{n_0} = 
 - i \int_{\mathcal{C},  \lambda[t_i] \to \lambda[t_f]}     \langle \phi_{n_0}(\lambda ) | d_\lambda   \phi_{n_0}(\lambda ) \rangle , 
\end{equation}
 where we used differential form notations. 
 Hence this Berry phase does not depends of the rate of variation of the parameters $\lambda(t)$ provided the condition of evolution within a single energy subspace
  is fulfilled : its contribution originates solely from the path $\mathcal{C}$ along which the systems evolves in parameters space. 

 By analogy with the Aharonov-Bohm  electromagnetic contribution (\ref{eq:AB}), we introduce a quantity playing the role of an electromagnetic potential, 
 and characterizing the source of Berry phase evolution at each point $\lambda$ of parameter space, defined as the $1$-form 
\begin{equation}
A_{n_0} (\lambda) = 
\frac{1}{i}   \langle \phi_{n_0}(\lambda ) | d_\lambda   \phi_{n_0}(\lambda ) \rangle 
=  \frac{1}{i} \sum_j \langle \phi_{n_0}(\lambda ) | \partial_{\lambda_j}   \phi_{n_0}(\lambda ) \rangle d\lambda_j . 
\label{eq:BerryConnexion}
\end{equation}
 Note that the normalization of the states  $|\phi_{n}(\lambda )\rangle$ ensures that $A$ is a purely real form, as it should be to enforce that 
 $\theta_{n_0} $ is a real phase. 
 Moreover, the Berry phase picked up by an eigenstate in an evolution along a closed loop in parameter space can be deduced from a quantity denoted as the Berry curvature $F(\lambda)$ analogous to the magnetic field, defined by the differential $2$-form : 
 \begin{align}
\theta_{n_0}   & = 
 \oint_{\mathcal{C}, \lambda[t_f] = \lambda[t_i]}     \langle \phi_{n_0}(\lambda ) |d_\lambda   \phi_{n_0}(\lambda ) \rangle  \\
 & =  \int_{\mathcal{S}, \partial \mathcal{S} = \mathcal{C}} d A(\lambda)  
    =  \int_{\mathcal{S}, \partial \mathcal{S} = \mathcal{C}} F  \textrm{ with } F = dA, 
\label{eq:BerryCurvature}
\end{align}
where $\mathcal{S}$ is a surface in parameter space whose boundary corresponds to $\mathcal{C}$. 
 A priori, the expressions (\ref{eq:BerryConnexion},\ref{eq:BerryCurvature}) appear to depend on the initial choice of continuous basis 
 $\{|\phi_n(\lambda)\rangle\}_n$. Let us 
 consider a second choice of continuous basis of eigenstates $\{|\phi'_n(\lambda)\rangle\}_n$. The assumption that the state $|\phi_{n_0}(\lambda)\rangle$ 
 is non degenerate implies that 
 $|\phi'_{n_0}(\lambda)\rangle=e^{if_{n_0}(\lambda)} |\phi_{n_0}(\lambda) \rangle$ where $f_{n_0}(\lambda)$ is continuous. Upon this change of  basis, the Berry connexion $A_{n_0}(\lambda)$ is modified 
 according to
\begin{equation}
A_{n_0}(\lambda) \to A'_{n_0}(\lambda)  = A_{n_0}(\lambda) + d f_{n_0} (\lambda)  . 
\end{equation}
 Hence the $1$-form $A_{n_0}((\lambda)$ depends on the exact reference basis used to transport eigenstates as $\lambda$ evolves, however, the associated Berry curvature $F_{n_0}$ does not, and is an intrinsic property of the space of eigenstates $\phi_{n_0}(\lambda)$. 

By inserting a completeness relation into the definition (\ref{eq:BerryConnexion}), we obtain the following alternative expression for the Berry curvature : 
\begin{equation}
 F_{n_0} (\lambda)  = - 
\sum_{m\neq n_0} 
\frac{
\langle \phi_{n_0}(\lambda ) | \partial_{\lambda_j}   \phi_{m}(\lambda ) \rangle ~
  \langle \phi_{m}(\lambda ) | \partial_{\lambda_l}   \phi_{n_0}(\lambda ) \rangle
}{
(E_{n_0}(\lambda) - E_m(\lambda))^2
} 
~
d\lambda_j \wedge d\lambda_l  . 
\label{eq:BerryCurvature2}
\end{equation}
A direct consequence of this second expression is that  the Berry curvature  summed over all eigenstates vanishes
 for any value of the parameters $\lambda$ :  
\begin{equation}
\sum_n F_n (\lambda) = 0 . 
 \end{equation}
 In the expression (\ref{eq:BerryCurvature2})  the Berry curvature depends on an apparent coupling term between energy bands $E_m(y)$, while 
 the initial expression (\ref{eq:BerryCurvature}) depends solely on properties of a given band. This second equation expresses in a manifest manner 
 the contraints imposed on the Berry curvature by the projection of the evolution onto the vector space of a single eigenstate. The closer the energies of other states, 
 the stronger this constraint.

\subsection{Berry phase and parallel transport on vector bundles}
\label{sec:BerryBundle}

The tools introduced above to describe the evolution of eigenstates of a Hamiltonian as a function of external parameters 
admit a natural interpretation in terms of parallel transport in vector bundle \cite{Simon1983}. Indeed, in this context 
 the Berry connexion had already been introduced by D. Thouless {\it et al.} \cite{Thouless:1982} to 
  characterize the topological properties of the quantum Hall effect on a lattice. We will naturally use this formalism in the following. 
Mathematically, the object we consider is a vector bundle, which generalizes the notion of tangent spaces over a manifold \cite{Nakahara}. 
 Such a vector bundle is constituted of a based space : at each point of this base space is associated a vector space called the fiber. In the present case 
 the base space is constituted of the manifold $\Lambda$ of external 
 parameters\footnote{Note that the Berry connexion can also be introduced on the projective vector bundle of the
                                   initial Hilbert space \cite{Aharonov:1987,Anandan:1990}.} 
 $\lambda$, the complex vector fiber is the Hilbert space~$\bf{H}$ of the problem, independent of $\lambda$.  
  The evolution with time $t$ of the parameters corresponds to a curve $\lambda(t), t\in[t_i,t_f]$ on the base space. 
   A vector $|\psi (\lambda) \rangle$ continuously defined as a function of $\lambda$ is called a section of the vector bundle.   
  To describe the evolution of vector of the fibers along a curve $\lambda(t)$, we need a prescription called a connexion or covariant derivative which generalizes 
  the notion of differential. Such a connexion defines how vectors are transported parallel to the base space. 
  In the present case, such a prescription is canonically defined : the fiber~$\bf{H}$ is independent of the point $\lambda$ considered and the fiber bundle can be written
as $\Lambda \times \bf{H}$. A natural choice to define parallel transport  of vectors amounts to choose a fixed basis $\left\{ |e_\mu \rangle \right\}_\mu$ of $\bf{H}$,
independent of the point $\lambda$. A vector is then parallel transported if it possesses fixed components in this basis : this amounts to define a covariant derivative or
 connexion $\nabla$ acting on sections and defined by 
\begin{equation}
|\psi (\lambda )  \rangle = f^\alpha(\lambda) | e_\alpha \rangle \rightarrow \nabla |\psi \rangle = (d f^\alpha) | e_\alpha \rangle . 
\label{eq:FlatConnexion}
\end{equation}    
The covariant derivative when acting on a vector tangent to the base space 
measures the "correction" necessary to the section  $|\psi (\lambda )  \rangle$ to transport it parallel to the frame 
$\left\{ |e_\mu \rangle \right\}_\mu$ in the direction of the vector. 
Hence parallel transported vectors corresponds to a choice of section satisfying   $\nabla |\psi (\lambda) \rangle = 0$. 

\begin{figure}[htbp!]
\begin{center}
\includegraphics[width=7cm]{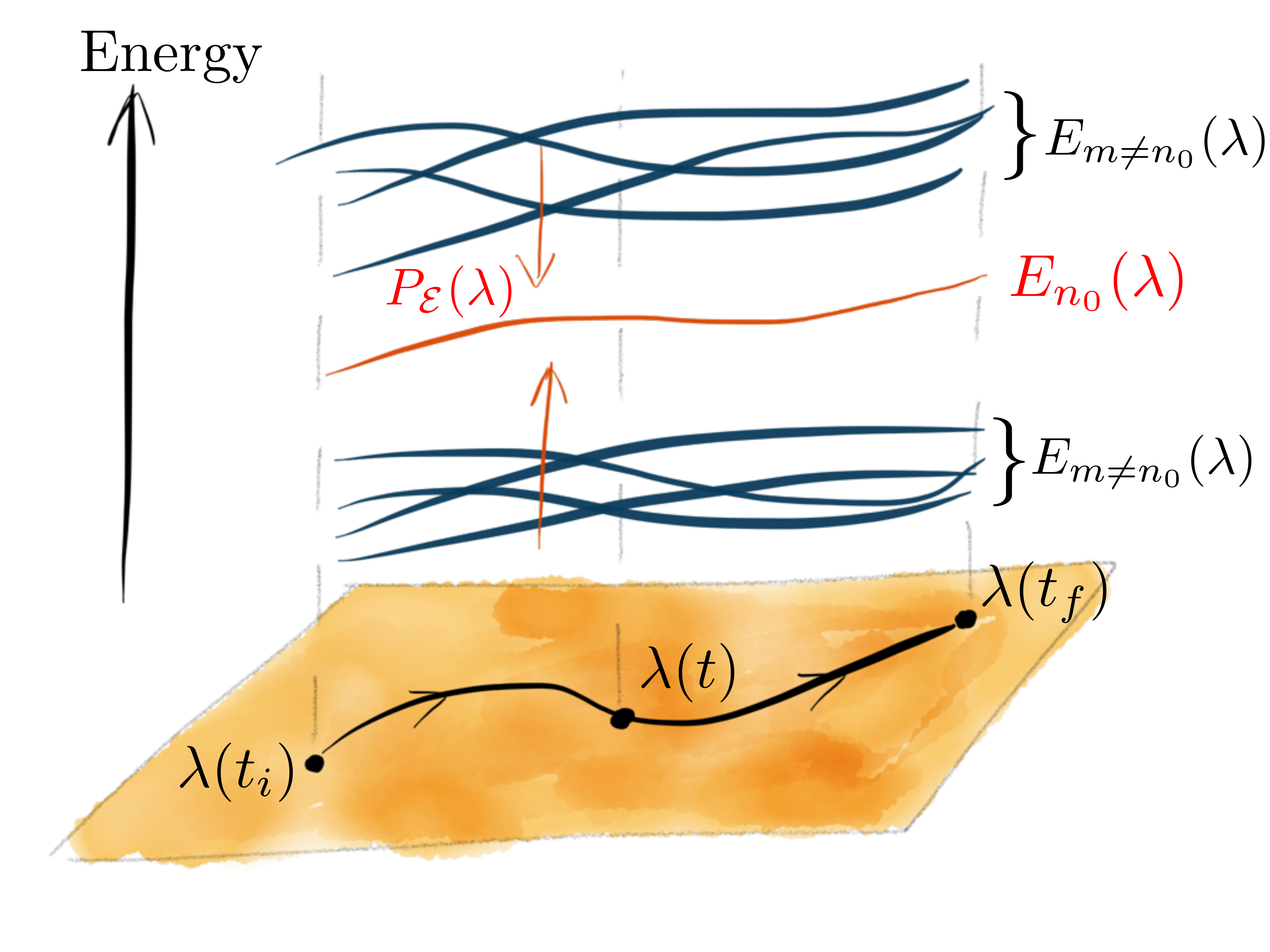}
\includegraphics[width=7cm]{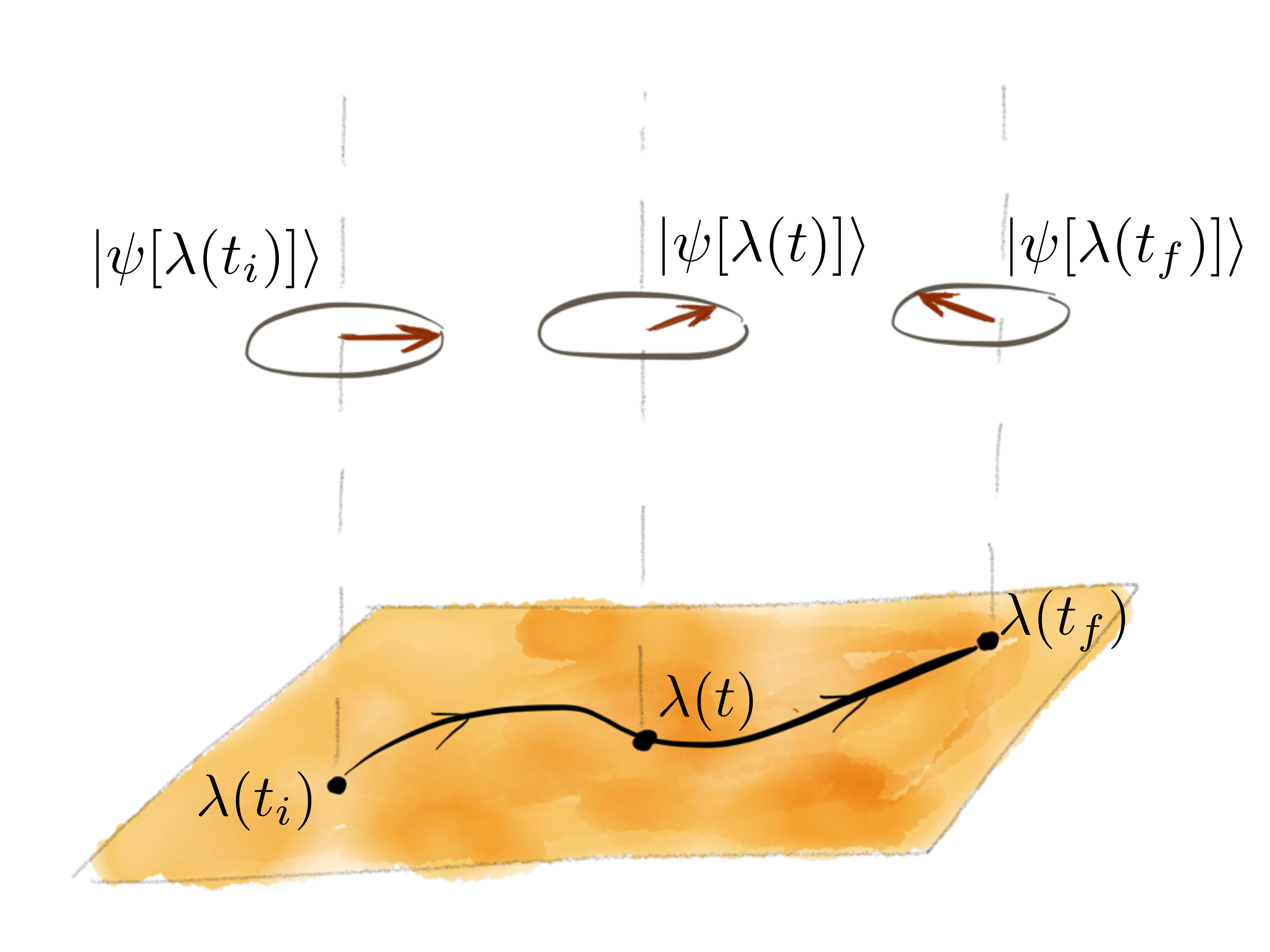}
\end{center}
\caption{Illustration the adiabatic evolution of an eigenstate as a function of external parameters $\lambda$ as the parallel transport of eigenstates over the manifold of parameters. }
\label{fig:Berry-1}
\end{figure}

 The problem considered in the previous section corresponds to an evolution with $\lambda$ restricted to a subspace of each fiber : the subspace $\mathcal{E}$ 
 associated with an energy eigenvalue $E_{n_0}(\lambda)$ (see Fig.~\ref{fig:Berry-1}). It is assumed that this energy remains separated from the rest of the spectrum by a spectral gap as $\lambda$ is varied. This allows to define unambiguously a projector onto this subspace 
$ P_{\mathcal{E}}(\lambda) =  |   \psi_{n_0} (\lambda ) \rangle \langle \psi_{n_0} (\lambda ) | $ where $| \psi_{n_0} (\lambda ) \rangle$ is a normalized eigenstate. 
 This subspace is constituted of a vector space of rank $1$ for each $\lambda$ 
 : if $|   \psi_{n_0} (\lambda ) \rangle$ is an eigenstate of $H(\lambda )$, so is 
 $ a | \psi_{n_0} (\lambda ) \rangle$ where $a \in \mathbb{C}$. The transport in this subspace is naturally defined by the 
 Berry connexion obtained by the projection in the subspace of the canonical connexion 
\begin{equation}
\Bnabla{\mathcal{E}} =  P_{\mathcal{E}} \nabla . 
\label{eq:DefBerryConnexion}
\end{equation}   
 A connexion is said to be flat if the parallel transport along a loop $\mathcal{C}$  which can be continuously deformed to a point 
 brings  any vector back to itself. This is obviously the case for the connexion $\nabla$ on the whole vector bundle, but not necessarily for the Berry connexion 
 $\Bnabla{\mathcal{E}}$. 
  As we will see in the following, only special bundles carry flat connections : this property is associated with a trivialization which is a family of smooth sections
$\lambda \mapsto |e^\alpha(\lambda ) \rangle $, defined over the whole base space and which form an orthonormal basis of every vector fiber.
 The flatness of a bundle can be locally probed through a curvature $2$-form $F$
 analogous to the gaussian curvature for manifold, and defined as 
 $\nabla^2  |\psi (\lambda )  \rangle = F  |\psi (\lambda )  \rangle$ for any section $|\psi (\lambda )  \rangle$. The property of the differential 
 immediately implies that of the curvature of the connexion (\ref{eq:FlatConnexion}) vanishes everywhere, as expected for flat connexion. 
 However, this is not the case for the Berry connexion. 

To express the curvature form of a Berry connexion, let us consider again a smooth section $|\psi (\lambda )  \rangle = f^\alpha(\lambda) | e_\alpha \rangle$. 
The Berry connexion acting on this section can be expressed as 
$\Bnabla{\mathcal{E}} |\psi (\lambda )  \rangle = \sum_\beta \langle e_\beta | \Bnabla{\mathcal{E}}  \psi (\lambda ) ~   | e_\beta \rangle $. 
 By using  $\nabla f^\alpha(\lambda) | e_\alpha \rangle = d f^\alpha(\lambda) | e_\alpha \rangle  $  we obtain 
\begin{equation}
\langle e_\beta | \Bnabla{\mathcal{E}} |\psi (\lambda ) \rangle  = 
\langle e_\beta | P_{\mathcal{E}} \nabla \left(  f^\alpha(\lambda) | e_\alpha \rangle  \right) 
=  d f^\alpha(\lambda)
 P^{\beta \alpha}  , 
\end{equation}
with $ P^{\beta \alpha} = \langle e_\beta | P_{\mathcal{E}}  | e_\alpha \rangle$.  
 The curvature of the Berry connexion, or Berry curvature, is  2-form curvature tensor  (analogous to the curvature tensor in the tangent plane of manifolds), 
 whose coefficients 
  \begin{equation}
 F^{\alpha \beta}_{{\mathcal{E}}}(\lambda) = \langle e_\alpha | \left( \Bnabla{\mathcal{E}} \right) ^2 |e_\beta \rangle  
\end{equation}
are expressed as 
 \begin{equation}
 F^{\alpha \beta}_{\mathcal{E}}(\lambda)  = \left( P_{\mathcal{E}}~ dP_{\mathcal{E}} \wedge dP_{\mathcal{E}} \right)^{\alpha \beta} = 
  P_{\mathcal{E}}^{\alpha \gamma} d P_{\mathcal{E}}^{\gamma \delta} \wedge dP_{\mathcal{E}}^{\delta \beta} , 
 \label{eq:BerryCurvatureTensor}
\end{equation}
which is often written in a compact form $F_{\mathcal{E}} = P_{\mathcal{E}}~dP_{\mathcal{E}} \wedge dP_{\mathcal{E}}$. 
Its trace $F = \sum_\alpha F^{\alpha \alpha}$ is denoted the scalar curvature 
(analogous to the gaussian curvature in the tangent plane of manifolds). 
 To make contact with the initial definition (\ref{eq:BerryConnexion},\ref{eq:BerryCurvature}) of Berry connexion and curvature, let us consider
the local frame of the bundle initially introduced and consisting of a basis of eigenstates $\{ |\phi_n(\lambda)\rangle\}_n$ of the Hamiltonian $H(\lambda)$. 
We consider the projector $P_{\mathcal{E}}$ onto the subspace of eigenvectors of energy $E_{n_0}(\lambda)$. By writing 
$ |\phi_{n_0}(\lambda)\rangle = f_{n_0}^\alpha (\lambda) |e_\alpha \rangle $ we recover the expressions of  Berry connexion and curvature 
$A(\lambda) = \langle \phi_{n_0} (\lambda) | \Bnabla{\mathcal{E}} | \phi_{n_0} (\lambda) \rangle 
=  \overline{f_{n_0}^\alpha} (\lambda) d  f_{n_0}^\alpha (\lambda) $ 
and $F = dA$. 
Having established the basic definitions of parallel transport of eigenstates, we can now turn ourselves to its translation in the context of energy bands in solids.

\subsection{Bloch Vector Bundle}
\label{sec:BlochFiber}

In section \ref{sec:Bloch}, we have established that the Bloch theory of electronic states in crystals amounts to identify  
an ensemble of $\alpha=1,\cdots,M$ energy bands $\epsilon^\alpha_k$ as well a quasi-periodic Bloch functions $\psi^{\alpha}_k$. 
 In doing so, we have replaced the initial Hilbert space $\bf{H}$ with  the collection of  complex vector spaces $\mathcal{H}_k$ 
 for $k$ in the Brillouin torus. Mathematically, this collection of vector spaces defines  a vector bundle $\mathcal{H}$ over 
the Brillouin torus $\BZ$, analogous to the vector bundle considered in the context of the Berry phase in the previous section. 
 It is thus tempting to translate the notion of Berry connexion and curvature in the context of the band theory of solids. Indeed, such a relation was established 
  soon 
 after the work of M. Berry by B. Simon \cite{Simon1983}. Several electronic properties of solids have now been related to these quantities, whose description goes 
 beyond the scope of the present paper (see  \cite{XiaoChangNiu2010} for a recent review). We will focus on the definitions necessary to the characterization of
  the topological properties of isolated bands in solids, as well as the specificities of parallel transport in electronic bands in solids, following \cite{Fruchart:2014a}. 
 
The first specificity of the Bloch bundle with respect to the example of section \ref{sec:BerryBundle} lies in the definition of fibers $\mathcal{H}_k$ which 
are vector spaces defined specifically at each point $k$, and thus different at different points $k$ and $k'$. 
 In the definition of the Berry connexion, we started from a flat connexion over the whole vector bundle, projected onto a vector space of eigenfunctions 
 (see eq.~(\ref{eq:DefBerryConnexion})). 
Defining a flat connexion amounts to identify a (local) trivialization of the bundle, {\it i.e.} a frame of sections $|e^{\un }_\alpha (k) \rangle$ providing a basis of 
each fiber $\mathcal{H}_k$. 
Not all bundles support such a trivialization globally: only those which are called trivial as we will see in the next section.  This is indeed the case for the Bloch bundle. However, this property is not obvious from the start : no canonical trivialization exist, as opposed to the situation considered for the Berry connexion where the fiber 
$\bf{H}$ was independent from the point $\lambda$ in base space. 
 Two such trivializations, also called Bloch conventions (see \cite{BenaMontambaux2009} for a discussion in graphene), are detailed in appendix \ref{sec:appendix}, 
 following  \cite{Fruchart:2014a}.
 The first one amounts to consider the Fourier transform of a basis of functions on a unit cell $\mathcal{F}$, typically localized 
 $\delta$ functions.  
These Fourier transform provide a trivialization of the Bloch bundle : a smooth set of sections which constitute a basis of each fiber $\mathcal{H}_k$. 
The associated connexion $\nabla^{\un}$ as well as the associated Berry connexions and curvatures generically  depends on the initial choice of fundamental domain 
$\mathcal{F}$. 
This choice of trivialization is useful when eigenstates and the Bloch Hamiltonian $H(k)$ are required to be periodic on the Brillouin torus, 
  {\it e.g.} in the geometrically interpretation of the topological properties of bands 
  (see {\it e.g.} \cite{Fruchart:2013}) or when studying  symmetry properties \cite{piechon:2014}. 

 A second choice of trivialization leads to a more canonical connexion: it corresponds to the 
 common writing of the Bloch functions $|\varphi (k) \rangle \in \mathcal{H}_k$ in terms of  functions $u(k,x)$ periodic on the crystal (with the periodicity of the Bravais lattice $\Gamma$) : 
\begin{equation}
\varphi (k;x)=\ee^{-\ii k\cdot(x-x_0)}u (k;x) .
\label{eq:PF2}
\end{equation}
A basis of periodic functions on the crystal $\mathcal{C}$, indexed by the sub lattices, is then ``pulled back" according to eq.~(\ref{eq:PF2}) 
as a trivialization of the Bloch bundle.  
This connexion only depends on the arbitrary choice of the origin  of space $x_0$ in (\ref{eq:PF2}). 
 If we choose another origin $x'_0$ then  $e'^{\deux}_\alpha (k;x) =\ee^{\ii k\cdot(x'_0-x_0)} e^{\deux}_\alpha (k;x)$ so that the two connexion differ by a closed one form : 
 $\nabla'^{\deux}=\nabla^{\deux}-\ii\,dk\cdot(x_0'-x_0)$.  
Note that as this connection $\nabla^{\deux}$ is flat (with a vanishing curvature $(\nabla^{\deux})^2=0$) 
 the associated parallel transport of any state along a contractible loop brings it back to itself. 
However, this is not so for the loops winding around the Brillouin torus : as can be seen from the relations (\ref{eq:PF},\ref{eq:ActionGamma*}). The 
connexion $\nabla^{\deux}$ is said to have non-trivial holonomies along these loops. 
 This second connexion $\nabla^{\deux}$ is particularly relevant for semi-classical analysis as it is related to the position operator by Fourier transform \cite{Blount1962} : 
\begin{equation}
(x-x_0)_j \psi(x)   \leftrightarrow -\ii~\nabla^\deux_j  \widehat{\psi}(k;x)  . 
\end{equation}
 Hence  physical quantities related to properties of a Berry connexion in crystals are indeed related to the properties of the Berry connexion associated with this second connexion $\nabla^{\deux}$.

\begin{figure}[thb]
\begin{center}
\includegraphics[width=14cm]{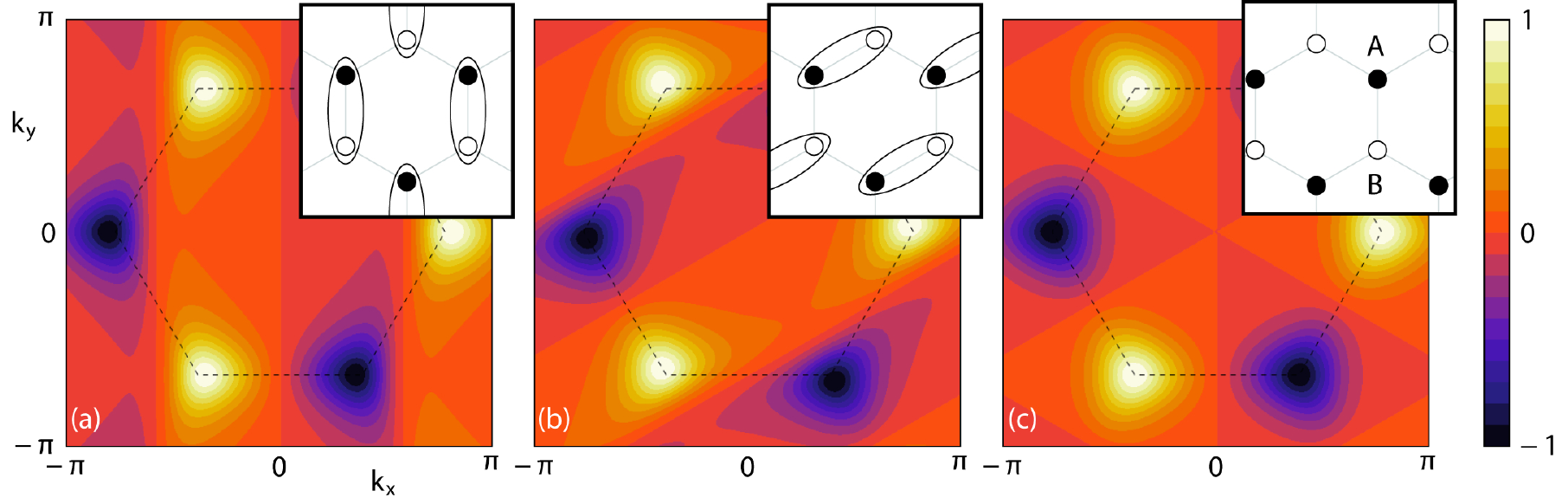}
\end{center}
\caption{The (normalized) Berry curvature $F(k)$ of the valence band 
in a gapped graphene model,  plotted on the Brillouin zone (dashed hexagone) for conventions I (a and b) 
and II (c). The corresponding choices of unit cells for convention I are shown in the insets. 
In all three cases, curvature $F(k)$ is concentrated around the Dirac points of ungapped graphene. It depends strongly on
the unit cell for convention I and is uniquely defined and respects the
symmetries of the crystal for convention II. After \cite{Fruchart:2014a}.}
\label{fig:BerryCurvature}
\end{figure}

 The necessity to distinguish these two natural connexions $\nabla^\un$ and $\nabla^\deux$ lies in their Berry connexions and curvatures, which
 generically differ on crystals with different lattices \cite{Fruchart:2014a}. This is in sharp contrast with the situation considered previously in 
 section \ref{sec:BerryBundle} where the Berry curvature was found to be uniquely defined.  This difference finds its origin in the different nature of the Bloch and Berry 
 bundle. As we will see in the following, topological properties of bands (sub-bundles) 
 do not depend on the choice  of flat connexion used initially. However, when studying geometrical Berry properties of these bands and their relevance in physics observables, one should pay special attention to be consistent in the use of connexion on the Bloch bundle. 
  To be more precise on the non unicity of Berry curvature, 
  let us consider two different trivializations  $\{ | e_\alpha^{(A)}\rangle \}_\alpha$ and $\{ | e_\alpha^{(B)}\rangle \}_\alpha$ of the Bloch bundle, 
 which can be {\it e.g.} 
 (i) a choice of type $\un$ and $\deux$ defined above,  
 (ii) two trivializations of types $\un$ corresponding to different choices of unit cells $\mathcal{F}$, 
 (iii) two trivializations of types $\deux$ with different choices of origin $x_0$. 
  Out of these two trivializations we can define two flat connexions $\nabla^{(A)}$ and $\nabla^{(B)}$ following (\ref{eq:FlatConnexion}), 
  and two Berry connexions $\Bnabla{\mathcal{E}}^{(A)}$ and $\Bnabla{\mathcal{E}}^{(B)}$ when projected 
  on the same sub-space $\mathcal{E}$  according to (\ref{eq:DefBerryConnexion}). Let us now express the corresponding  Berry connexion tensors  
 (\ref{eq:BerryCurvatureTensor}) in the same trivialization and compare the corresponding scalar curvature (their trace) $F^{(A)}$ and $F^{(B)}$. 
  The change of basis defines a local unitary matrix $U$ by 
 $ | e_\alpha^{(A)}\rangle=U^{\beta}_\alpha (k)  | e_\beta^{(B)}\rangle$. The two projection matrices corresponding to $P_\mathcal{E}$ are related
 through 
 $P^{(B)}(k)=U(k).P^{(A)}(k).U^{-1}(k)$. The two Berry curvatures are found to differ by  \cite{Fruchart:2014a}
\begin{equation}
	F^{(B)}=F^{(A)}+{\rm tr}\left(dP^{(A)}(k)\wedge U^{-1}dU-P^{(A)}(k)\,U^{-1}dU\wedge U^{-1}dU \right).
\label{eq:VariationCurvature}
\end{equation}
 As a consequence, generically operators $U$ and $P$ do not commute : we obtain 
 \begin{itemize}
 \item[-] a Berry  curvature $F^{\un}$ which depends on the choice of unit cell $\mathcal{F}$, and  differs from the  Berry  curvature $F^{\deux}$,
 \item[-]  a uniquely defined quasi-canonical Berry  curvature $F^{\deux}$, which is independent of the choice of origin $x_0$. Indeed a change of origin corresponds to 
scalar matrix $U$ which commutes with any projector.
\end{itemize} 
 This dependance of Berry curvatures on the Bloch convention or trivialization was noted in 
 \cite{FuchsPiechonGoerbigMontambaux2010} in the context of graphene, and is 
 illustrated on Fig.~\ref{fig:BerryCurvature} : the value of this Berry curvature for the valence band of a model of gapped graphene (model of Boron Nitride) is illustrated for two different choices of unit cell in convention $\un$ and for convention $\deux$. Details of the analysis can be found in  \cite{Fruchart:2014a}. 

\section{Topological Properties of a Valence Band in an Insulator}

\begin{figure}[thb]
\begin{center}
\includegraphics[width=5cm]{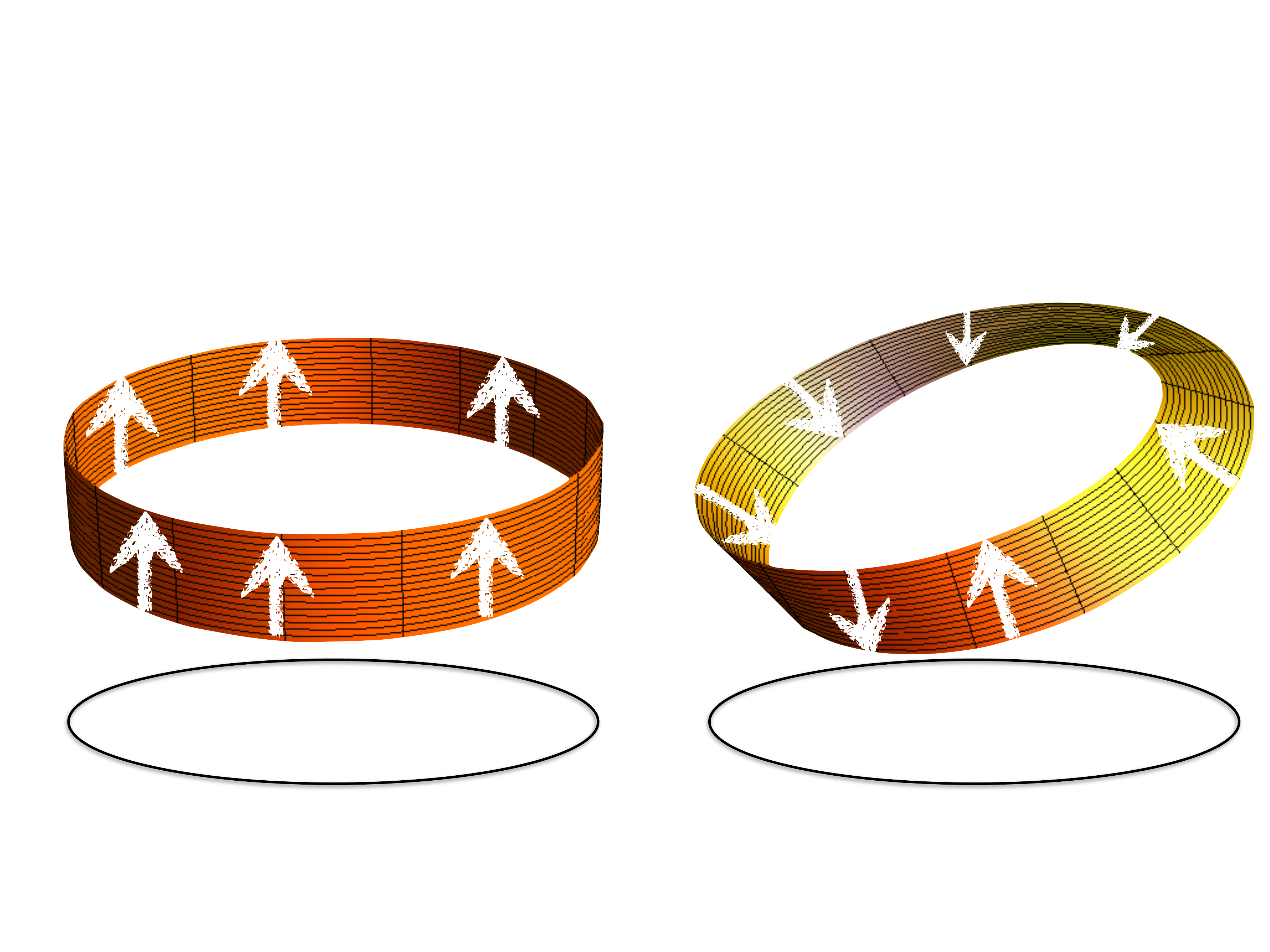}
\includegraphics[width=4.5cm]{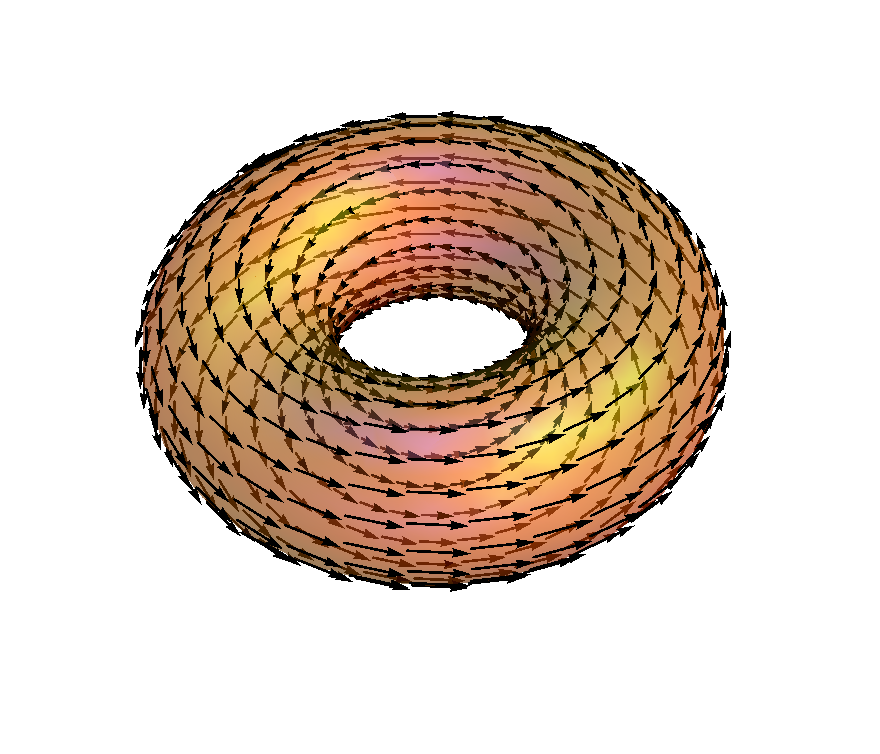}
\includegraphics[width=4.5cm]{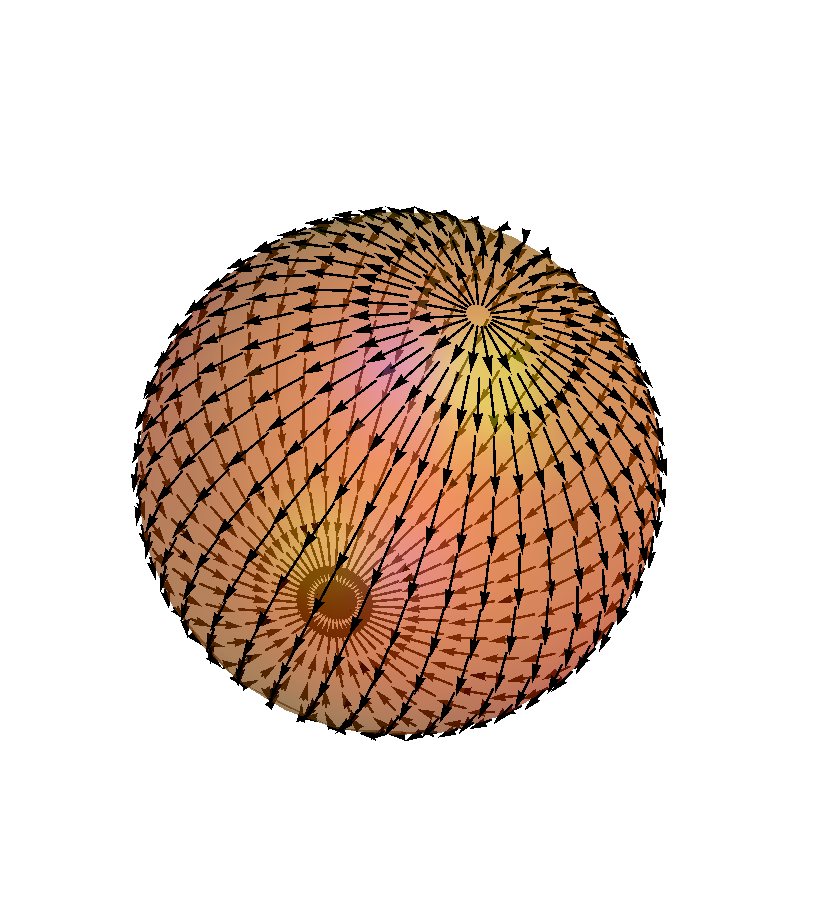}
\end{center}
\caption{Illustration of topology of vector bundles. Left : a trivial (the cylinder) and a non trivial (the M{\"o}bius strip) vector bundle defined as real bundle on the circle $\mathcal{C}_1$. Right : trivial and non trivial bundles of tangent vectors on the two dimensional torus and the sphere. }
\label{fig:TopologyBundle}
\end{figure}

We can now discuss the notion of topological properties of a band in a crystal, or vector bundle. 
A fiber bundle is said to be topologically trivial if it can be deformed into the product of its base space times a fixed vector space : this corresponds to the situation encountered in the previous section of bundle supporting a  set of basis of the fibers continuous on the whole base space, called a frame of sections. Such a bundle, called trivializable,  
was shown to support a flat connexion, {\it i.e.} a connexion whose curvature vanishes everywhere. 
 Such a continuous basis is not supported by all vector bundles : when this is not the case, the bundle is said to be topologically non-trivial. By definition, on a 
 non trivial vector bundle 
 non-vanishing  vector fields (set of sections) do not exist: they necessarily possess  singularities. We will use this property in the 
 following to detect a non trivial topology of bands in the Bloch bundle. 
Two examples of non trivial vector bundles are represented in figure~\ref{fig:TopologyBundle} : the M{\"o}bius strip, a real bundle on the circle contrasted with the 
trivial cylinder $\mathcal{S}_1\times \mathbb{R}$, and the problem of
 a real vector field tangent to the sphere (the hairy ball problem).  
In both cases, a continuous vector field defined on the bundle necessarily possesses a singularity. 

 From the discussion in the previous section, we know that the Bloch bundle is trivial : it supports continuous basis of the fibers. Non trivial topology can only arise when
 it is decomposed into two different sub-ensembles (sub-bundles) which can themselves be non-trivial. This is indeed the case in an insulator : the gap splits the spectrum into valence and conduction bands. This allows to split each Bloch space $\mathcal{H}_k$ into the vector space of valence band states and the 
 vector space of conduction band states whose corresponding vector bundles can then acquire topological properties defining a topological insulator. 
 We will see in the following section two different types of non trivial topology that can arise in such a valence bundle : the first one associated with the
  Chern index when no symmetry is present, and the second one associated with the Kane-Mele index when time-reversal symmetry is enforced. 
For simplicity, we will discuss these properties in two dimensions, on the Haldane and Kane-Mele models. Note that while the Chern index only characterizes 
two dimensional insulators, the Kane-Mele index can be generalized to three dimensions.

\subsection{Chern topological insulators}
\label{sec:ChernInsulator}

\subsubsection{Two Band Insulator}
\label{sec:TwoBandModel}

The simplest insulator we can imagine consists of two bands : one above and one below the band gap. The description of such a simple insulator is 
provided by a $2 \times 2$ Bloch Hamiltonian $H(k)$ at each point $k$ of the Brillouin zone and parameterized by the real functions $h_{\mu}(k)$ according to  
\begin{equation}
H(k) = h^{\mu}(k) \sigma_{\mu} = h_0(k) \sigma_0 + \vec{h}(k) \cdot \vec{\sigma{}}
\label{eq:GenericTwoLevelHamiltonian}
\end{equation}
in the basis of  $\sigma_{0} =  \mathds{1}$ and Pauli matrices $\sigma_\mu$. 
This is the Hamiltonian of a spin $\frac12$ coupled to a magnetic field which depends on point $k$. 
If we choose a trivialization of type $\un$ of the Bloch bundle (basis of Bloch spaces $\mathcal{H}_k$), see section \ref{sec:BlochFiber}, 
the real  functions $h_{\mu}$ are periodic on the Brillouin torus. 
The  two energy bands, defined as eigenvalues of $H(k)$,  are  
\begin{equation}
\epsilon_{\pm}(k) = h_{0}(k) \pm h(k)
\end{equation}
with $h(k)= |\vec{h}(k)|$. An insulator situation implies that the two bands never touch each other : therefore $h(k)$ should never vanish on the whole Brillouin torus 
for the gap to remain finite. We will focus on this situation. The energy shift $h_0$ of both energies has no effect on topological properties we will discuss: for simplicity 
we also set  $h_0 = 0$.
 If we use spherical coordinates to parametrize $\vec{h}(k)$, 
the eigenvectors of the valence band $\epsilon_-(k)$ below the gap are defined, up to a phase, by
\begin{equation}
    \vec{h}(h(k),\phi(k),\theta(k)) = h \, \begin{pmatrix}
        \sin \theta \, \cos \varphi \\
        \sin \theta \, \sin \varphi \\
        \cos \theta
    \end{pmatrix}
\quad
\longrightarrow 
\quad
| u_{-}(\vec{h}(k)) \rangle = 
\begin{pmatrix}
- \sin \frac{\theta}{2} \\ 
\ee^{\ii \varphi} \, \cos \frac{\theta}{2}
\end{pmatrix} 
\label{eq:FilledEigenvectorObstruction}
\end{equation}
From this expression, we realize that the norm $h=|\vec{h}|$ does not affect the eigenvector $| u_{-}(\vec{h}(k)) \rangle$ : therefore, the parameter space describing the vectors of the valence band is 
 a $2$-sphere $S^2$. A continuous vector field $| u_-\rangle$ cannot be defined  on $S^2$ : it possesses necessarily singularities (vortices).  
 In other words the corresponding vector bundle on the sphere $S^2$ is not trivial : the valence band vector bundle on the Brillouin torus is then nontrivial as soon as  
 the function $k \mapsto \vec{h}(k)$ covers the whole sphere.  This behavior unveils the nontrivial topology of a vector bundle on the sphere, discovered by Dirac and Hopf in 1931 \cite{Dirac1931}. It is related to the well known property that the phase of the electronic state around a magnetic monopole cannot be continuously defined : 
 it is necessarily singular at a point, here the North pole $\theta=0$ where the state  (\ref{eq:FilledEigenvectorObstruction}) still 
 depends on the ill-defined angle $\phi$. 
 Indeed the situation where $\vec{h}(k)$ covers the whole sphere as $k$ varies in the Brillouin torus necessarily corresponds to a situation, if $\vec{h}(k)$ was a magnetic field, where the interior of the torus would contain a monopole. 
 This situation exactly describes a non-trivializable vector bundle : if $\vec{h}(k)$ covers the whole sphere, 
 whatever the change of phase we try on the state  (\ref{eq:FilledEigenvectorObstruction}), it is impossible 
 to find a continuous basis of the valence band bundle defined everywhere.

  How to detect in general that the ensemble of eigenstates of bands below the gap possess such a topological property ? If  $\vec{h}(k)$ was 
 a magnetic field, we would simply integrate its magnetic flux through the Brillouin torus to detect the magnetic monopole. For more general vector bundles, 
 the quantity playing the role of the magnetic flux is the Berry curvature introduced in section \ref{sec:BerryBundle} : 
 it is the curvature associated with any Berry connexion. When integrated over the Brillouin zone (the base space of the vector bundle), it defines the so-called first Chern number 
\begin{equation}
c_{1} = \frac{1}{2 \pi} \, \int_{\text{BZ}} \ F .
\label{eq:ChernNumberDefinition}
\end{equation}
This integral of a $2$-form is only defined on a $2$-dimensional surface : this Chern number characterizes bands of insulators in dimension $d=2$ only.
 Let us also notice that this Chern number, which is a topological property of the bundle, is independent of the connexion chosen. We have seen in 
 section \ref{sec:BlochFiber} that different Berry connexions with different Berry curvatures could be defined on a Bloch sub bundle : 
 they all give rise to the same value of the integral in eq.~(\ref{eq:ChernNumberDefinition}) as can be checked from 
eq.~(\ref{eq:VariationCurvature}). 

Let us now calculate this Chern number for the valence band of the model \eqref{eq:GenericTwoLevelHamiltonian} : 
 the curvature 2-form takes the form \cite{Berry1984}
\begin{equation}
F = \frac{1}{4} \, \epsilon^{i j k} \, \frac{1}{h^{3}} \, h_{i} \, \dd h_{j} \wedge \dd h_{k} 
=
\frac{1}{2}
\,
\frac{\vec{h}}{|{h}|^3}
\cdot
\left(
\frac{\partial \vec{h}}{\partial k_x}
\times   
\frac{\partial \vec{h}}{\partial k_y}
\right)
\; \dd k_{x} \wedge \dd k_{y} , 
\label{eq:BerryCurvatureTwoLevelSystem}
\end{equation}
corresponding to a Chern number 
\begin{equation}
c_{1}
= \frac{1}{4 \pi}  \int_{\text{BZ}}
\,
\frac{\vec{h}}{|h|^3}
\cdot
\left(
\frac{\partial \vec{h}}{\partial k_x}
\times\frac{\partial \vec{h}}{\partial k_y}
\right)
\; \dd k_{x} \wedge \dd k_{y}
\label{eq:ChernNumberAsIntegral}
\end{equation}

One recognises in the expression \eqref{eq:ChernNumberAsIntegral} 
the index of the function $\vec{h}$ from the Brillouin torus to the sphere $S^2$, which exactly counts the "winding" of this map around the sphere, as expected. 

\subsubsection{The Haldane Model}

 We can apply this characterization of valence bands to the study of phases of a toy-model defined on graphene, and first proposed by D.  Haldane \cite{Haldane88}. 
The corresponding first quantized Hamiltonian can be written as:
\begin{equation}
 H = 
    t \, \sum_{\langle \, i, j \,  \rangle} |i\rangle \langle j | 
    +
    t_{2} \, \sum_{\langle \langle \, i, j \, \rangle \rangle} |i\rangle \langle j | 
    +
    M \, \left[
    \sum_{i \in A} |i\rangle \langle i | 
    -
    \sum_{j \in B}|j\rangle \langle j | 
    \right]
\label{eq:HaldaneHamiltonianFirstQuantized}
\end{equation}
where $|i\rangle$ represents an electronic state localized at site $i$ (atomic orbital), 
$\langle \, i, j \, \rangle$ represents nearest neighbors  sites and 
$\langle \langle \, i, j \, \rangle \rangle$  second nearest neighbors sites
 on the honeycomb lattice (see Fig.~\ref{fig:HoneycombLattice}), while 
$i \in A$ corresponds to a sum over sites in the sublattice $A$ (resp. $i \in B$ in the sublattice $B$).
This Hamiltonian is composed of a first nearest neighbors hopping term with a hopping amplitude~$t$, 
a second neighbors hopping term with a hopping parameter $t_{2}$, and a last sublattice symmetry breaking  term 
with on-site energies $+M$ for sites of sublattice $A$, and $-M$ for sublattice $B$,  
which thus breaks inversion symmetry. 
Moreover, each unit cell of the is supposed to be threaded by a magnetic fluxes which averaged to zero, but lead to  Aharanov--Bohm phases 
which break time-reversal symmetry, and  are taken into account through  the Peierls substitution 
$  t_{2} \to t_{2} \ee^{\ii \phi}$ while $t$ is independant of the Aharonov--Bohm phase $\phi$. 
The Fourier transform (according to convention $\un$ defined previously) of the Hamiltonian \eqref{eq:HaldaneHamiltonianFirstQuantized} 
leads to a $2 \times 2$ Bloch Hamiltonian 
$
H(k) = h^{\mu}(k) \sigma_\mu
$
in the $(A,B)$ sublattices basis with 
\begin{subequations}
\label{eq:Haldane_h}
\begin{align}
h_{0} &= 2 t_{2} \cos \phi \sum_{i=1}^{3} \cos( k \cdot b_{i} ) 
\quad ; \quad 
h_{z} = M - 2 t_{2} \sin \phi \sum_{i=1}^{3} \sin( k \cdot b_{i} ) 
\quad ;
\label{eq:Haldane_hz}
\\
h_{x} &= t \left[1 + \cos(k \cdot b_{1}) + \cos(k \cdot b_{2}) \right] 
\quad ; \quad 
h_{y} = t \left[\sin(k \cdot b_{1}) - \sin(k \cdot b_{2}) \right] 
\quad ; 
\label{eq:Haldane_hy}
\end{align}
\end{subequations}
with a convention where $\vec{h}$ is periodic: $\vec{h}(k + G )=\vec{h}(k)$ and the vectors $b_i$ are defined in figure \ref{fig:HoneycombLattice}. 

\begin{figure}[th!]
\centering
\includegraphics[width=8cm]{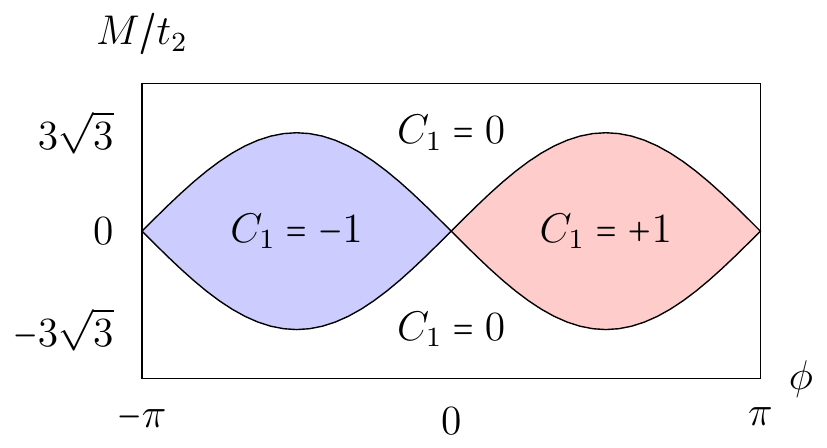}   
\caption{Phase diagram of the Haldane model (\ref{eq:Haldane_h}) as a function of the Aharonov-Bohm flux $\phi$ (amplitude of time-reversal breaking term) 
and alternate potential $M/t_{2}$ measuring the inversion breaking amplitude. 
Insulating phases are characterized by the Chern number $c_{1}$.}
\label{fig:HaldanePhaseDiagram}
\end{figure}

To determine the phase diagram, we first consider the energy difference $|h(k)|$ between the two bands which is positive for all points $k$ in the Brillouin zone except
 for $|M|/t_2 = 3 \sqrt{3}  \sin \phi$ where it vanishes at one point in the Brillouin zone, and two points only for the case of graphene $M=\phi=0$. 
 Hence we obtain the phase diagram of Fig.~\ref{fig:HaldanePhaseDiagram}  which consist of three  gapped phases (insulators) separated by transition lines. 
  Along these critical lines,  the system is not insulating anymore: it is a semi-metal with low energy Dirac states. 
  The insulating phases are not all equivalent : their valence bands possess different topological Chern number. As a consequence, the transition between these different insulating phases necessarily occurs through a gap closing transition. These Chern number can be determined directly using the expression 
 (\ref{eq:ChernNumberAsIntegral}), or in a simpler manner by realizing that the winding of the function $\vec{h}(k)$ around the sphere also correspond to 
 the algebraic number it will cross a ray originating from the center of the sphere (as $\vec{h}(k)$ describes a closed surface $\Sigma$ on the sphere)  \cite{SticletPiechonFuchsKukuginSimon2012}. 
 By choosing such a line as the $Oz$ axis, we identify the wavevectors $k$ such that $\vec{h}$ crosses the $z$ axis corresponding to 
 $h_{x}(k)=h_{y}(k)=0$ : they correspond to the Dirac points $\vec{K},\vec{K}'$ of graphene. 
The Chern number is then expressed as  
\begin{equation}
c_{1} = \frac{1}{2} \, \sum_{k =K,K'} \text{ sign } \left[ 
h(k) \cdot n(k)
\right], 
\end{equation}
where $n(k)=\pm \hat{e}_{z}$ is the normal vector to $\Sigma$ at $k$. We obtain 
\begin{equation}
c_{1} = \frac{1}{2} \left[
 \text{ sign } \left( \frac{M}{t_{2}} + 3 \, \sqrt{3} \sin(\phi) \right)
 - 
 \text{ sign } \left( \frac{M}{t_{2}} - 3 \, \sqrt{3} \sin(\phi) \right)
 \right], 
\end{equation}
which corresponds to the original result of \cite{Haldane88} : the corresponding values for the gapped phases are shown on Fig.~\ref{fig:HaldanePhaseDiagram}).

\subsection{Kane-Mele topological index with time-reversal symmetry}

\subsubsection{Time-reversal symmetry in Bloch bands}

\begin{figure}
\centering
\includegraphics[width=6cm]{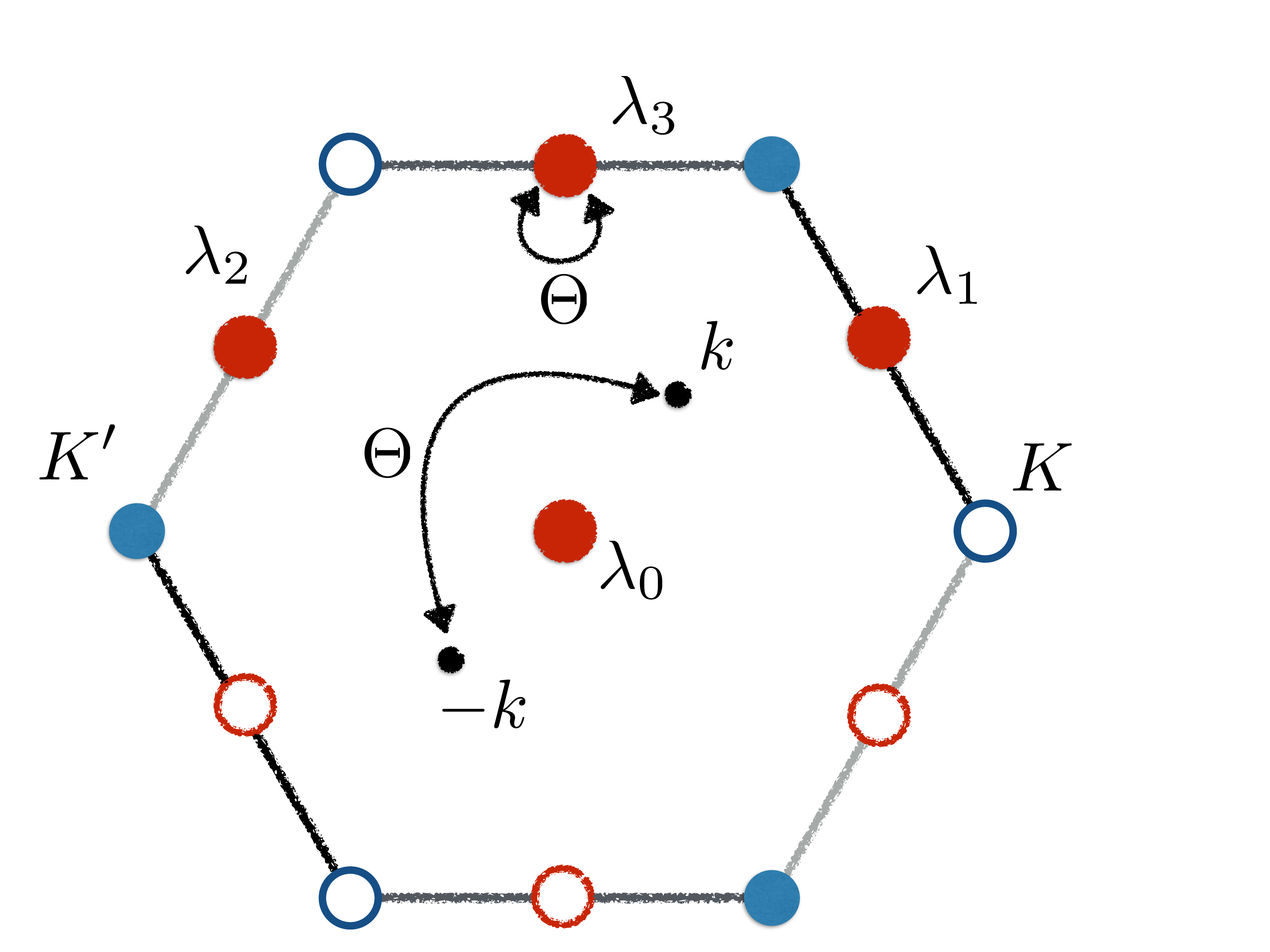}      
\caption{Representation of the Brillouin torus of graphene (triangular Bravais lattice) with the $K$ and $K'$ Dirac points, 
the points $k$ and -$k~(+G)$ related by time reversal operator $\Theta$ and 
the  four time-reversal invariant momenta 
$\lambda_i$. 
}
\label{fig:TRIM2D}
\end{figure}

Within quantum mechanics, the time-reversal operation is described by an anti-unitary operator 
$\Theta$ \cite{SakuraiModernQM}, which satisfies  
$\Theta(\alpha x) = \overline{\alpha} \Theta(x)$ for $\alpha \in \mathbb{C}$ and    $\Theta^\dagger = \Theta^{-1}$.
 When spin degrees of freedom are included it can be written as  a $\pi$ rotation in the spin space $\Theta = \ee^{- \ii \pi S_{y} / \hbar } \; K$, 
where $S_{y}$ is the $y$ component of the spin operator, and $K$ is the complex conjugation. 
Therefore for half-integer spin particles time-reversal operation posses the crucial property: $\Theta^2 = - \mathds{I}$. This property has important consequences
 for the band theory of electrons in crystals. 
In this context, as time-reversal operation relates momentum $k$ to $-k$, $\Theta$ acts as an anti-unitary operator from the Bloch fiber $\mathcal{H}_k$  to 
$\mathcal{H}_{-k}$. Time-reversal invariance of a band structure implies that the Bloch Hamiltonians at $k$ and $-k$ satisfy:
\begin{equation}
H(-k) = \Theta H(k) \Theta^{-1}
\label{eq:TimeReversalBlochHamiltonian}. 
\end{equation}
This equation implies that if $|\psi(k)\rangle$ is any eigenstate of the Bloch Hamiltonian $H(k)$ then $\Theta |\psi(k)\rangle$ 
is an eigenstate of the Bloch Hamiltonian $H(-k)$ at $-k$, with the same energy. This is the Kramers theorem. 
Moreover $\Theta^2 = - \mathds{I}$ implies that these two  Kramers partners are orthogonal.
 In general the two Kramers partners lie in different Bloch fibers $\mathcal{H}_k$  and $\mathcal{H}_{-k}$, except at specials points where 
 $k$ and $-k$ are identical on the torus, {\it i.e.} differ by a reciprocal lattice vector  $k=-k+G \Leftrightarrow k=G/2$. 
These points are called time-reversal invariant momenta (TRIM), and denoted by $\lambda_i$ in the following. 
 They are represented on the Fig.~\ref{fig:TRIM2D} for the Brillouin torus of graphene. 
Above these points, 
the two Kramers partners live in the same fiber $\mathcal{H}_{\lambda_i}$ and the spectrum is necessarily degenerate. 

\begin{figure}
\centering
\includegraphics[width=6cm]{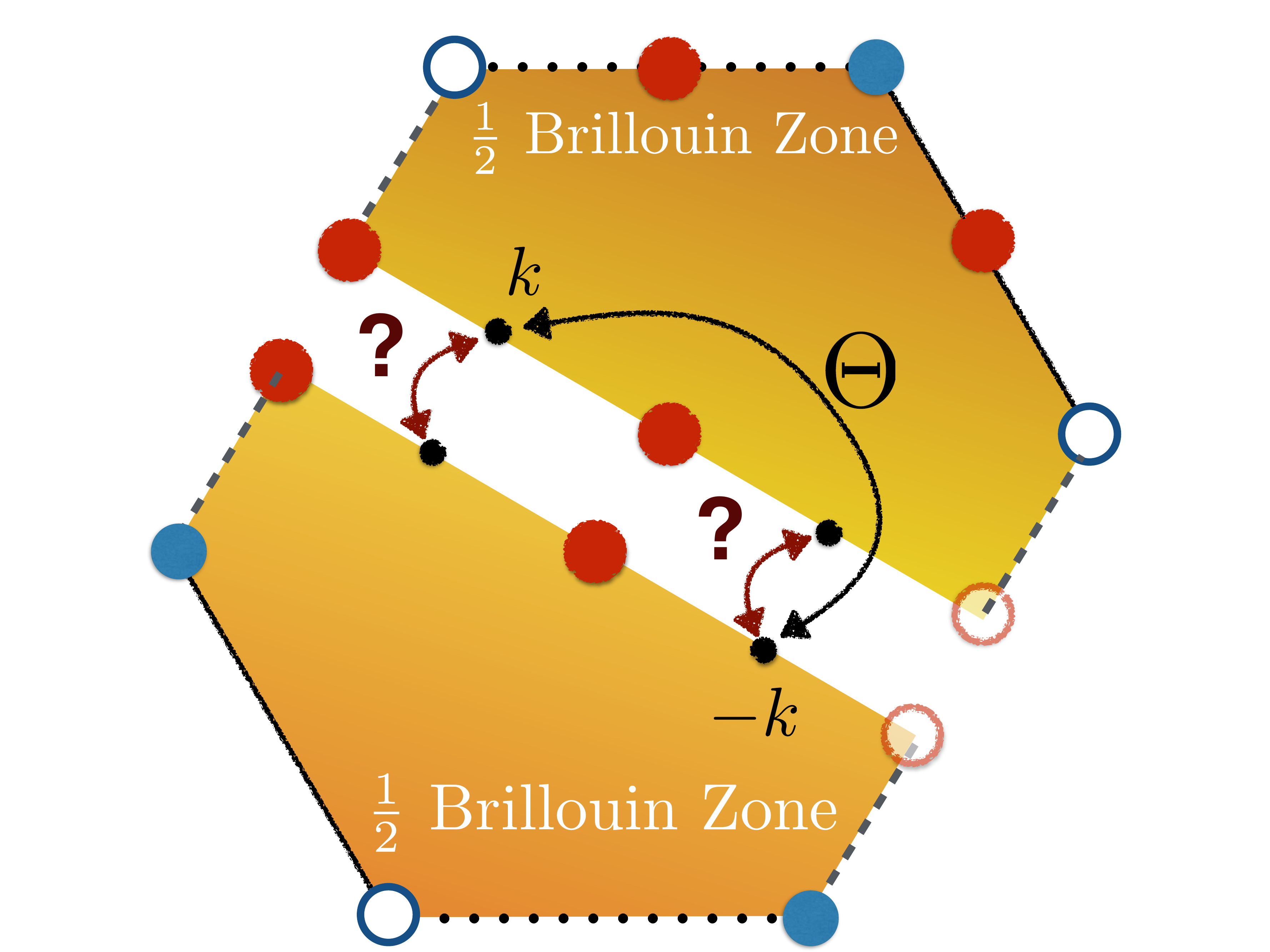}      
\caption{Eigenstates of an isolated band (such as valence bands in an insulator) can be defined continuously on half  of the Brillouin Zone. When time reversal symmetry is present, the eigenstates on the other half are defined unambiguously by the action of the time reversal operator $\Theta$. The smooth reconnection between the eigenstates of the two halves is not guaranteed :  a  winding of the relative phase between the two opposite states prevents such a possibility. The Kane-Mele 
$\mathbb{Z}_2$ topological index probes this possible obstruction to define continuously eigenstates over the Brillouin zone as a consequence 
of time-reversal symmetry constraints.  
}
\label{fig:KaneMeleObstruction}
\end{figure}

Note that in a time-reversal invariant system of spin $\frac12$ particles, the Berry curvature within valence bands is odd: $F_\alpha(k)=-F_\alpha(-k)$. Hence 
 the Chern number of the corresponding bands $\alpha$ vanishes: the valence vector bundle is always trivial from the point of 
 view of the first Chern index. 
 It is only when the constraints imposed by time-reversal symmetry are considered that a different kind of non-trivial  topology  can emerge. 
 The origin of this new topological index can be understood as follows : to enforce the constraints imposed by time reversal symmetry, it appears advisable to define 
 the pairs of spin $\frac12$ eigenstates on half of the Brillouin torus. It is always possible to do so in a continuous manner. By application of the time reversal operator 
 $\Theta$, a continuous pair of eigenstates is then defined on the second half of the Brillouin torus. However, a smooth reconnection of the eigenstates at the boundary of the two halves Brillouin torus is not guaranteed and not always possible up to a deformation of the initial eigenstates : see Fig.~\ref{fig:KaneMeleObstruction}. 
 The new Kane-Mele topological index  aims precisely at measuring this possible obstruction to define continuously Kramers pairs on the whole Brillouin torus. It is intimately related to the constraints imposed by time-reversal symmetry : it will be topological in the sense that the associated properties are robust with respect to all perturbations that preserve this symmetry and the isolation of the band considered. Similar topological indices associated with stronger crystalline symmetries can be defined, whose discussion goes beyond the scope of this paper.

\subsubsection{Kane-Mele  model}
\label{sec:Z2:BHZLikeModel}

 For simplicity, we will restrict ourselves to the discussion of the new $\mathbb{Z}_2$ index in two dimensions, although it can be generalized to $d=3$ as opposed to the Chern index. 
We will start with a discussion of the model initially proposed by Kane and Mele \cite{KaneMele2005}, in a slightly simplified form 
\cite{FuKane2007}. This model describes a sheet of graphene in the presence of a particular form of spin-orbit coupling. While the proposed Quantum Spin Hall topological phase has not been determined in graphene due to the small amplitude of spin-orbit in Carbon atoms, it constitutes the simplest model to discuss this new topological index, and shares a lot of common features with the Haldane model of section \ref{sec:ChernInsulator}, which justifies our choice to discuss it.

 By Kramers theorem, the simplest  insulator with spin-dependent time-reversal symmetric band structure 
   possesses two bands of Kramers doublet below the gap, and two bands above the gap : it is described by a 
  $4 \times 4$ Bloch Hamiltonian. While such a Hamiltonian is generally parametrized by $5$ $\Gamma_a$ matrices 
 which satisfy a Clifford algebra $\{ \Gamma_{a} , \Gamma_{b}\} = 2 \delta_{a,b}$ and their commutators  \cite{FuKane2007}, a minimal model 
 sufficient for our purpose can be written in terms of three such matrices  \cite{KaneMele2005,FuKane2007}
\begin{equation}
H(k) = 
d_{1}(k) ~\Gamma_{1}
+d_{2}(k) ~ \Gamma_{2}
+d_{3}(k) ~\Gamma_{3} . 
\label{eq:KM-Hamiltonian}
\end{equation}
These gamma matrices are constructed as tensor products of Pauli matrices that represent two two-level systems : in the present case they correspond to the
 sub lattices $A$ and $B$  of the honeycomb lattice (see Fig.~\ref{fig:HoneycombLattice}) and the spins of electrons. In the basis
\begin{equation}
\mathrm{(A, B)} \otimes (\uparrow, \downarrow) = \mathrm{(A  \uparrow, A  \downarrow, B  \uparrow, B  \downarrow)} , 
 \label{eq:TensorProductBasis}
 \end{equation}
these matrices read 
\begin{equation}
\Gamma_{1} = \sigma_{x} \otimes I 
\ ; \ 
\Gamma_{2} = \sigma_{y} \otimes I 
\ ; \ 
\Gamma_{3} = \sigma_{z} \otimes s_{z}.
\label{eq:defGamma}
\end{equation}
where  $\sigma_{i}$ and $s_{i}$ are sub lattice and spin Pauli matrices. 
 The first two terms are spin independent and describe nearest neighbor hopping of electrons in graphene.  
 $d_1(k),d_2(k)$ accounts for the hopping amplitudes on the honeycomb lattice (see {\it e.g.} eq.~(\ref{eq:Haldane_hy})). 
The only term acting on the spins of electrons is the last term, which plays the role of spin-orbit : here this choice of matrix is  dictated by simplicity such as to 
preserve the $s_z$ spin quantum numbers. This not necessary but simplifies the discussion. A general spin-orbit coupling will obviously not satisfy this constraint, but will not modify the topological properties we will identify provided it doesn't close the gap. 
With those choices, the time-reversal operator reads:
\begin{equation}
\Theta = \ii \, ( \mathds{I}  \otimes  s_{y} ) \, K. 
\end{equation}
while the parity (or inversion) operator exchanges the $A$ and $B$ sub lattices : 
\begin{equation}
P = \sigma_{x} \otimes \mathds{I} ,  
\end{equation}
whose two opposite eigenvalues correspond to the symmetric and antisymmetric combination of $A$ and $B$ components. 
Let us now enforce the time-reversal symmetry of this model : the commutations rule 
\begin{equation}
\Theta \Gamma_1 \Theta^{-1} = \Gamma_1
\ ; \ 
\Theta \Gamma_2 \Theta^{-1} = - \Gamma_2
\ ; \ 
\Theta \Gamma_3 \Theta^{-1} = - \Gamma_3
\end{equation}
imply that the constraint $\Theta H(k) \Theta^{-1} = H(-k)$ is fulfilled provided $d_1(k)$ is an even and $d_2(k),d_3(k)$ odd functions on the Brillouin torus : 
\begin{equation}
d_1(k) = d_1(-k +G)  \ ; \ 
d_2(k) = - d_2(-k +G)  \ ; \ 
d_3(k) = - d_3(-k +G)  \ ; \ 
\textrm{ for }G\in \Gamma^* .
\label{eq:oddness}
\end{equation}
 Note that with these constraints, the Bloch Hamiltonian also satisfy $P H(k) P^{-1} = H(-k)$ : parity is also a symmetry of the model, with eigenvalues $d_1(\lambda_i)$ at the TRIM points.  
 As a consequence, the $P\Theta$ symmetry of the model implies that the spectrum is degenerate over the whole Brillouin torus and not only at the TRIM 
 $\lambda_i$ points. 

Diagonalisation of the Hamiltonian (\ref{eq:KM-Hamiltonian},\ref{eq:defGamma}) yields the eigenenergies 
$\epsilon_\pm (k) = \pm d(k) $ with  $d(k) = (d_{1}^{2} + d_{2}^{2} + d_{3}^{2})^{1/2}$. Hence an insulator correspond to the situation where $d_1, d_2$ and $d_3$ cannot simultaneously vanish. We will restrict ourself to this situation. The eigenvectors for the valence band $\epsilon_-(k)$ below the gap can be determined 
 with special care to ensure Kramers degeneracy :
 \begin{equation}
|u_{-,\downarrow}\rangle = 
\frac{1}{\mathcal{N}_{1}} \;
\begin{pmatrix}
    0 \\
    - d_{3} - d \\
    0 \\
    d_{1} + \ii d_{2}
\end{pmatrix}
\qquad
\text{and}
\qquad
|u_{-,\uparrow}\rangle = 
\frac{1}{\mathcal{N}_{2}} \;
\begin{pmatrix}
    d_{3} - d \\
    0 \\
    d_{1} + \ii d_{2} \\
    0
\end{pmatrix}, 
\label{eq:Z2:GrapheneLike:Eigenstates}
\end{equation}
where $\mathcal{N}_{j}(\vec{d})$ are normalization factors. 
For these states, singularities appear when $d_{1}=d_{2}=0$. Through the polar decomposition $d_{1}+ \ii d_{2} = t \, \ee^{\ii \theta}$, we obtain in the limit $t \to 0$:
\begin{subequations}
\begin{equation}
|u_{-,\downarrow}\rangle \to 
\begin{pmatrix}
    0 \\
    - 1 \\
    0 \\
    0
\end{pmatrix}
\qquad
\text{and}
\qquad
|u_{-,\uparrow}\rangle  \to 
\begin{pmatrix}
    0 \\
    0 \\
    \ee^{\ii \theta} \\
    0
\end{pmatrix}
\qquad
\qquad
\text{ for $d_{3} > 0$, }
\label{eq:Z2:GrapheneLike:LimitOfEigenvectorsAtPositived5}
\end{equation}
\begin{equation}
|u_{-,\downarrow}\rangle \to 
\begin{pmatrix}
    0 \\
    0 \\
    0 \\
    \ee^{\ii \theta}
\end{pmatrix}
\qquad
\text{and}
\qquad
|u_{-,\uparrow}\rangle \to 
\begin{pmatrix}
    -1 \\
    0 \\
    0 \\
    0
\end{pmatrix}
\qquad
\qquad
\text{ for $d_{3} < 0$}.
\label{eq:Z2:GrapheneLike:LimitOfEigenvectorsAtNegatived5}
\end{equation}
\label{eq:Z2:GrapheneLike:LimitOfEigenvectors}
\end{subequations}
The phase $\theta$ is ill-defined when $t \to 0$, so one of these eigenstates is singular in this limit depending on the sign of $d_3$. 
 In the Brillouin zone, the Kramers theorem imposes that these singularities necessarily occur by pairs $\pm k_0$ of singularity of opposite vorticity for the phase
  ($d_3(k_0)$ and  $d_3(-k_0)$ possessing opposite signs). 
 Note that with our choice of eigenstates (\ref{eq:Z2:GrapheneLike:Eigenstates}), these singularities occur exactly at the Dirac points $\pm K$ 
of graphene, which are defined by $d_{1}(\pm K)=d_{2}(\pm K)=0$. This is a consequence of our choice of eigenstates with a well-defined 
$S_z$ spin quantum number. The reader can convince himself that when such a pair of singularity exist, its location $k_0$ can be modified by a 
general $U(2)$ rotation in the space of eigenvectors\footnote{%
I am indebted to Krzysztof Gawedzki for a clarification on this point.}, 
{\it i.e.} at the expense of preserving the $S_z$ spin quantum number for the eigenstates. 
(see also \cite{GrafPorta2012} for a related point of view). 

 Thus triviality of the valence band bundle depends on the existence in the Brillouin torus of points where $d_1$ and $d_2$ simultaneously vanish. Topologically, this existence depends solely on the sign of $d_1$ at the four TRIM points $\lambda_i$ : condition (\ref{eq:oddness}) imposes that 
 the odd functions $d_2(k),d_3(k)$ must vanish at the $\lambda_i$ points, but also along arbitrary lines connection these  $\lambda_i$ points. On the other hand 
 the  functions $d_1(k)$ is necessary non-vanishing at the $\lambda_i$ points. 
If the $d_1(\lambda_i)$ all have the same signs, {\it e.g.} positive, then a single choice of nonsingular eigenvectors $|u_{-,\uparrow \downarrow}(k)\rangle$ over the whole Brillouin zone is possible : the valence band bundle is trivial. On the other hand if one of the $d_1(\lambda_i)$ have a sign opposite to the other, $d_1(k)$ vanishes 
on align around this $\lambda_i$. Necessarily there exists then two points where both $d_1$ and $d_2$ vanish : the valence band bundle is twisted. 
If $d_1(\lambda_i)$ takes a different sign at a pair of TRIM, the singularities can be avoided and the valence band bundle is again trivial. 
  Through this reasoning, we realize that the triviality of the valence band bundle is directly related to the sign of 
\begin{equation}
\prod_{\lambda_i} \textrm{sign} \, d_{1}(\lambda_i) = (-1)^\nu
\label{eq:KaneMeleIndex}
\end{equation}
  When this product is positive, the valence band bundle is trivial, while it is not trivial when this product is negative. This defines a new $\mathbb{Z}_2$ 
  topological index $\nu$, called the  Kane-Mele index. 
 Morerover this expression can be easily generalized to three dimensional model as shown by Fu, Kane, and Mele \cite{FuKaneMele2007}. 
It is of practical importance to realize that in this expression, the $d_{1}(\lambda_i)  $ correspond to the parity eigenvalues of the band at the high symmetry points 
$\lambda_i$ : 
a simple recipe to induce a  $\mathbb{Z}_2$ topological order is to look for valence bands of an insulator where the valence bands possess a 
\emph{band inversion} around one of the symmetry point, {\it i.e.} where the natural order of atomic orbitals composing the bands has been reversed. A band 
inversion between two orbitals of opposite parity induces a topological twist of the valence bands. 

\begin{figure}
\centering
\includegraphics[width=8cm]{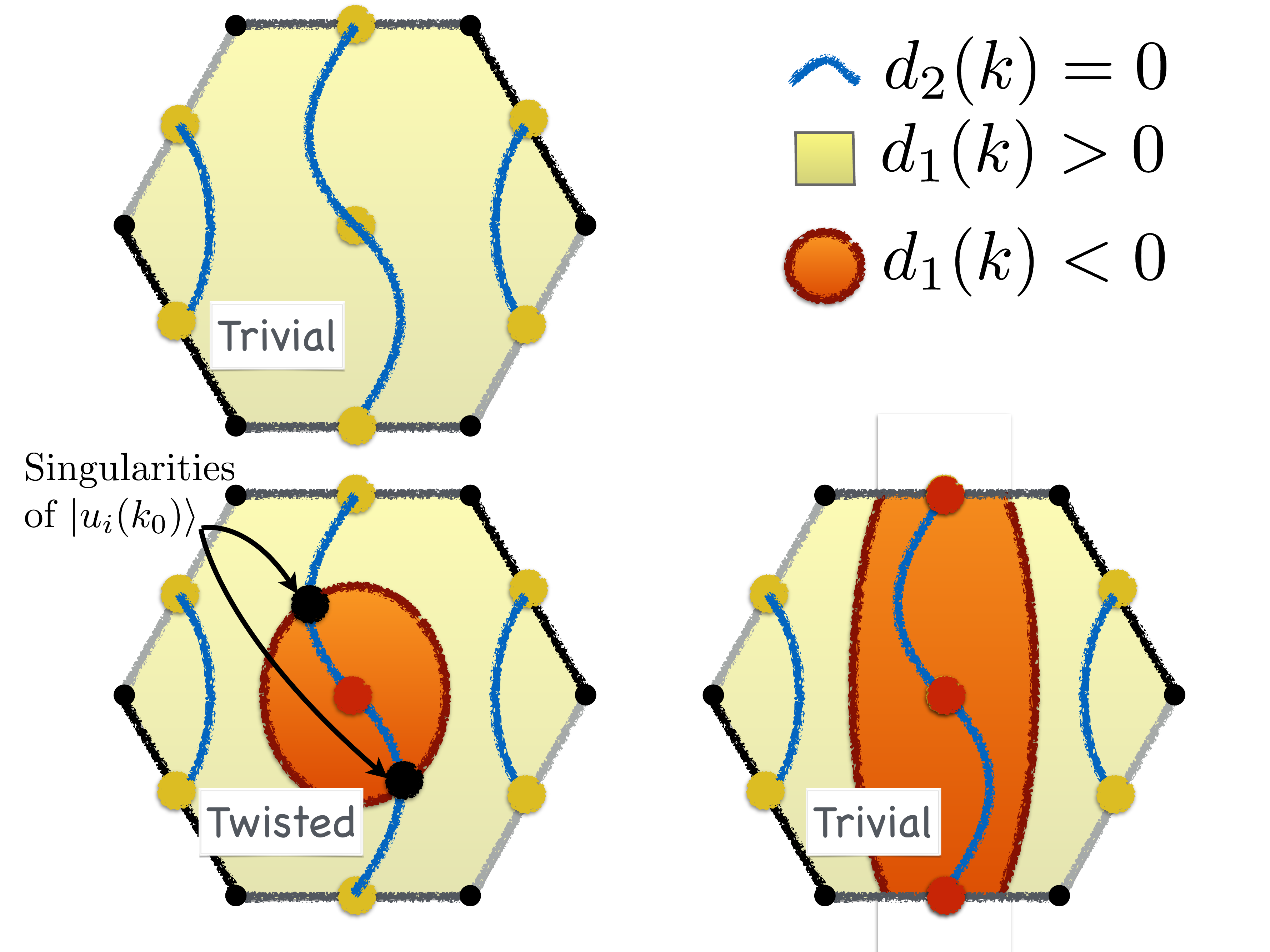}      
\caption{Illustration of the condition of existence of singularities for the valence band eigenfunctions $|u_-(k)\rangle$ of the Kane-Mele model. 
These singularities exist when $d_1(k)=d_2(k)=0$ which necessarily occurs when $\prod_i d_1(\lambda_i) <0$ : this is the condition which 
defines the topological non-triviality of these valence bands. 
 }
  \label{fig:KaneMeleVortices}
\end{figure}

\subsubsection{Expression of the Kane-Mele index}

\begin{figure}
\centering
\includegraphics[width=15cm]{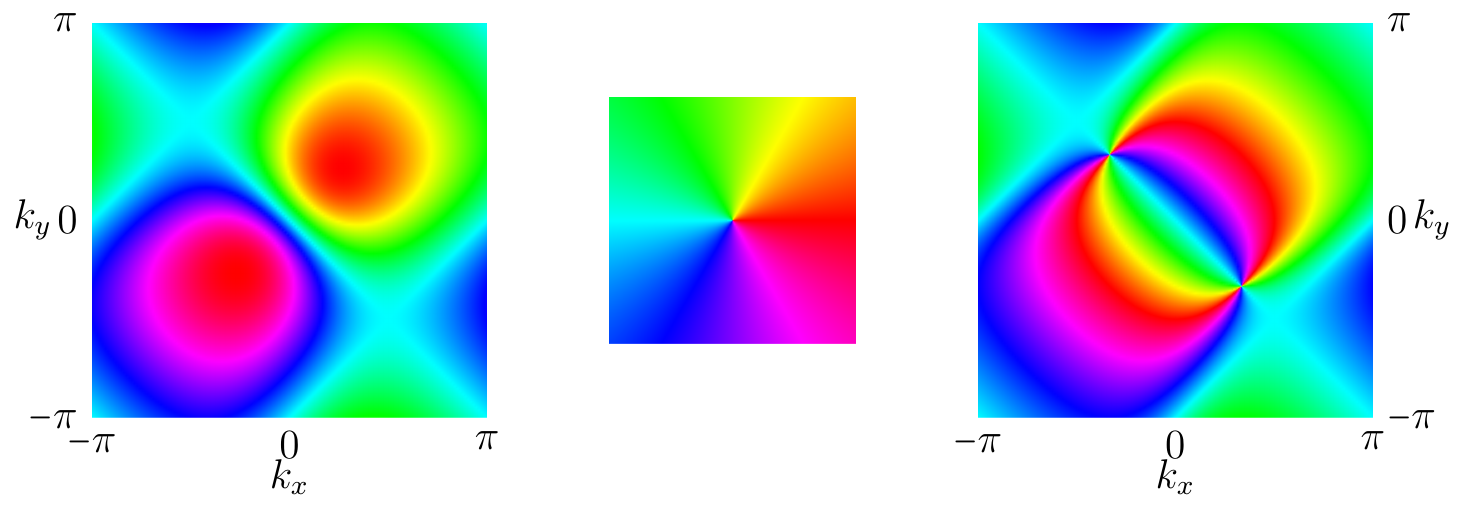}      
\caption{Phase of the pfaffian $\text{Pf } (m)$ on the Brillouin zone for the Kane-Mele model (\ref{eq:KM-Hamiltonian},\ref{eq:defGamma})
 in a trivial (left, $\mu=-3$) and topological insulating phases (right, $\mu=-1$). 
The parametrizing functions adapted from \cite{FuKane2007} are 
$d_{1}(k_x,k_y) = \mu + \cos k_x + \cos k_y$, $d_{2}(k_x,k_y) = \sin k_x + \sin k_y$, 
$d_{3}(k_x,k_y) = \sin k_x - \sin k_y - \sin (k_x-k_y)$. 
The middle figure represent the phase color convention. }
  \label{fig:PhasePfaffian}
\end{figure}

 The expression (\ref{eq:KaneMeleIndex}) for the so-called Kane-Mele index was derived on more rigorous ground for models with parity symmetry, while more 
 general expression exist when parity is absent. The original derivation proceeds as follows \cite{KaneMele2005}. 
 As the Chern number of the valence band vanishes, there 
 exist a trivialization of the corresponding bundle, {\it i.e.}  global continuous states  $\{ |e_\alpha (k)\rangle \}_{\alpha=1,\dots, 2m}$ constituting a  basis of 
 the filled band fiber  at each point $k$. 
Kane and Mele originally considered the matrix element of the time-reversal operator $\Theta$ in this basis\footnote{%
To define the scalar product of vectors between the different the fibers  at $k$ and $-k$, we use the trivialization as
$\textrm{BZ} \times \mathbb{C}^{2m}$ of the valence bundle  Bloch states and the scalar product in $\mathbb{C}^{2n}$.} \cite{KaneMele2005} : 
\begin{equation}
m_{\alpha \beta}(k) = \langle e_{\alpha}(k) | \Theta e_{\beta}(k) \rangle . 
\label{eq:KMTRMatrixDefinition}
\end{equation}
This $2m\times 2m$ matrix is not unitary, but it is antisymmetric (because $\Theta^{2} = -1$) : it possesses a well defined 
 Pfaffian $\text{Pf } (m)$.  
This quantity tracks the orthogonality between Kramers related eigenspaces of the valence band.  
This property can be monitored by considering the vortices of this Pfaffian. 
Time reversal invariance imposes that these vortices necessarily come by pair $k_0,-k_0$ of opposite vorticities. On the other hand, 
at any time-reversal invariant points $\lambda_i $, the Pfaffian has unit modulus $|\text{Pf } m(\lambda_i)| = 1$ and thus cannot support a vortex. 
 A continuous deformation of the vectors constituting the valence band allows to move continuously these vortices in the Brillouin zone, however two 
 situations occur : 
 \begin{itemize}
 \item[-] for an even number of vortices, vortices of opposite vorticity can  annihilate by pair away from the $\lambda_i$ points. Topologically this situation is equivalent to the absence of vortex. 
 \item[-]  for an odd number of vortices, one pair of opposite vortices remain, which cannot be annihilated through the $\lambda_i$ points. 
 \end{itemize}
 These considerations allowed Kane and Mele to define their index $\nu$ as the parity of the number of pairs of vortices of $\text{Pf } m(k)$ in the Brillouin zone, 
 or more practically the parity of the number of vortices in half of the Brillouin zone  : 
\begin{equation}
\nu =
 \frac{1}{2 \pi \ii} \; \oint_{\partial \frac12 \text{BZ}} \dd \log \left[ \text{Pf } (m) \right]   \quad \text{mod } 2 .
\label{eq:Z2InvariantPfaffian}
\end{equation}
As an illustration, let us apply this expression to the Kane-Mele model (\ref{eq:KM-Hamiltonian},\ref{eq:defGamma}).  In this case, the Pfaffian is: 
%
%
$\text{Pf } (m) =  d_{1} (d_{1} + \ii d_{2})/ ( (d_{1}^{2} + d_{2}^{2})^{1/2} d)$ 
which possesses vortices at the points $d_{1}=d_{2}=0$ where the valence band eigenstates were previously found to be singular. These vortices as represented in Fig.~\ref{fig:PhasePfaffian}. 
Alternate and more practical expressions of the  $\nu$ invariant exist: one of them, introduced by Fu and Kane \cite{FuKane2006}, 
uses the so-called sewing matrix 
\begin{equation}
w_{\alpha \beta}(k) = \langle e_{\alpha}(-k) | \Theta e_{\beta}(k)\rangle , 
\label{eq:SewingMatrixDefinition}
\end{equation}
to define the $\mathbb{Z}_2$ invariant as 
\begin{equation}
(-1)^{\nu}
=
\prod_{\lambda_i}
\frac{\textrm{Pf} ~w(\lambda_i)}{\sqrt{\det w(\lambda_i)}} , 
\label{eq:Z2InvariantSewingMatrix}
\end{equation}
which simplifies into 
\begin{equation}
(-1)^{\nu} = \prod_{\lambda_i } \textrm{sign} \, d_{1}(\lambda_i)
\end{equation}
when parity symmetry is present with eigenvalues $d_{1}(\lambda_i)$ at the TRIM. 
 A different expression formulates explicitly the invariant as an obstruction to defined continuously the phase of Kramers partners \cite{Roy2009, FuKane2006}. 
 This expression can also be defined from a homotopy point of view  allowing for an easy generalization to three dimension\cite{MooreBalents2007}. It allows for practical numerical determination of the Kane-Mele topological index \cite{SoluyanovVanderbilt2011}.
Other expressions of the Kane-Mele $\mathbb{Z}_2$ invariant have also been discussed in relation to a Chern-Simons topological effective field theory \cite{QiHughesZhang:2008,WangQiZhang:2010}.

%
%
%
%

\subsection{Insulating materials displaying a Topological Valence Band}

The Kane-Mele model discussed above is certainly oversimplified to allow a realistic description of materials possessing a Kane-Mele non-trivial topology. 
However, its study allowed to identify two crucial ingredients when seeking topological insulators~: 
\begin{itemize}
\item[-] we should look for materials with a strong spin dependent interaction preserving time-reversal symmetry, and spin-orbit is the obvious candidate. 
\item[-] a topologic order is associated with the inversion at one of the high symmetry point of the Brillouin zone of two bands of opposite parities. This implies that one of the bands (typically the highest) below the gap should originates around this point from atomic orbitals which usually have a higher energy. Such a \emph{band inversion} is typically associated with a strong spin-orbit coupling.  
\end{itemize}
 The band inversion by spin-orbit condition imposes to look for materials with a gap comparable with the weak amplitude of spin-orbit. 
Hence the materials in which this topological order has been tested or proposed  are small gap semi-conductors. 
Several of them such as HgTe/CdTe or InAs/GaSb  proposed by the group of S.-C. Zhang \cite{BHZ2006, Liu:2008} to realize the two dimensional topological phase 
being commonly used as infrared detectors with energy gaps below $1.5$ eV compatible with the infrared wavelengths. 
 It is important to realize that while the topological index probes a global property of eigenstates over the whole Brillouin zone, the crucial property of formula 
(\ref{eq:KaneMeleIndex}) is that the topological non-triviality can be detected locally at one of the symmetry points ! This is a remarkable property of practical importance. Indeed, most studies of topological insulators properties focuses on an effective $k.P$ description of eigenstates close the minimal gap \cite{Bastard} which is 
certainly not valid in the whole Brillouin zone, but yet capture all essential ingredients of these new phases. 
 Beyond the quantum wells built out of HgTe/CdTe or InAs/GaSb semiconductors already mentioned, and in which the two dimensional Quantum Spin Hall phase has been experimentally detected, 
canonical three dimensional topological insulators include the Bi compounds such as Bi$_2$Se$_3$, Bi$_2$Te$_3$, or bulk HgTe strained by a CdTe substrate.

\section{From Insulators to Semi-metals}

\subsection{From Topology in the Bulk to Semi-metallic Surface States}

The essential consequence of a nontrivial topology of valence bands in an insulator is the appearance of \emph{metallic edge states} at its surface, with a linear relativistic dispersion relation. 
It is these surface states that are probed experimentally when testing the topological nature of a semi-conductor, and which give rise to all physical properties of these 
materials. 
The continuous description of the interface between two insulators amounts to continuously deform the bulk eigenstates of one into the other 
(see \cite{Bastard}). If one of the two insulators is topological this is not possible for the ensemble of valence band eigenstates by definition of the topological nature of the twist. Extrapolation of wave functions from  both sides requires to close the gap : a smooth connection is then always possible as a consequence of
 the triviality of the Bloch bundle. 
Another more practical point of view consists in recalling that the $\mathbb{Z}_2$ topological order can be identified with the band inversion at one of 
the $\lambda_i$ TRIM point of the Brillouin zone with respect to the ``standard ordering" of atomic orbitals constituting bands in a standard insulator.   
Hence at the interface between two such insulators, the two corresponding bands have to be flipped back from the inverted to the normal ordering. 
When doing so, they necessarily give rise to energy bands within the gap and located at the interface, corresponding to edge or surface states. 
Moreover, the crossing of two such bands can be described at low energy by a linear dispersion relation, associated with a relativistic Dirac equation 
of propagation. 

 These edge states can be explicitly derived for the two two-dimensional models considered in this article : 
 let us first consider the interface at $y=0$ between two different insulating phases of 
 the Haldane model (\ref{eq:HaldaneHamiltonianFirstQuantized}). At a transition line, the gap  $m=h_{z}(K)= M - 3 \sqrt{3} t_{2} \sin \phi $ 
 at one of the two Dirac points of graphene  ($h_{x}(K)=h_{y}(K)=0$) changes sign. 
 Assuming that the Bloch hamiltonian is not modified significantly away from this point $K$  at the transition, we describe the interface by focusing solely on the low 
 energy degree of freedom and linearize the Bloch Hamiltonian around point $k=K+q$ to obtain a massive Dirac Hamiltonian with mass $m$ : 
\begin{equation}
H_{\text{l}}(q) =\hbar v_{\text{F}} \left( q_x \cdot \sigma_{x} +  q_y \cdot \sigma_{y}  \right) + m(y) \, \sigma_{z}, 
\label{eq:HI}
\end{equation}
with $m=h_{z}(K+q,y)$ such that $m(y<0)<0$ and $m(y>0)>0$. In real space, this Dirac equation reads (with $\hbar v_{\text{F}} = 1$) : 
\begin{equation}
H_{\text{l}} = - \ii ~\nabla \cdot \sigma_{\text{2d}} + m(y) \sigma_{z} . 
\end{equation}
Solving this equation, we obtain a state of energy $E(q_{x}) = E_{\text{F}} + \hbar v_{\text{F}} q_{x}$ localized at the interface $y=0$  \cite{JackiwRebbi1976}.   : 
\begin{equation}
\psi_{q_{x}}(x,y) \propto \ee^{\ii q_{x} \, x} \, \exp \left[ - \int_{0}^{y} m(y') \, \dd{}y' \right] \; 
\begin{pmatrix}
    1 \\
    1
\end{pmatrix} . 
\end{equation}
 This edge state has a positive group velocity $v_{\text{F}}$ and thus corresponds to a ``chiral right moving'' edge state
\begin{figure}[ht]
\centering  
\includegraphics[width=8cm]{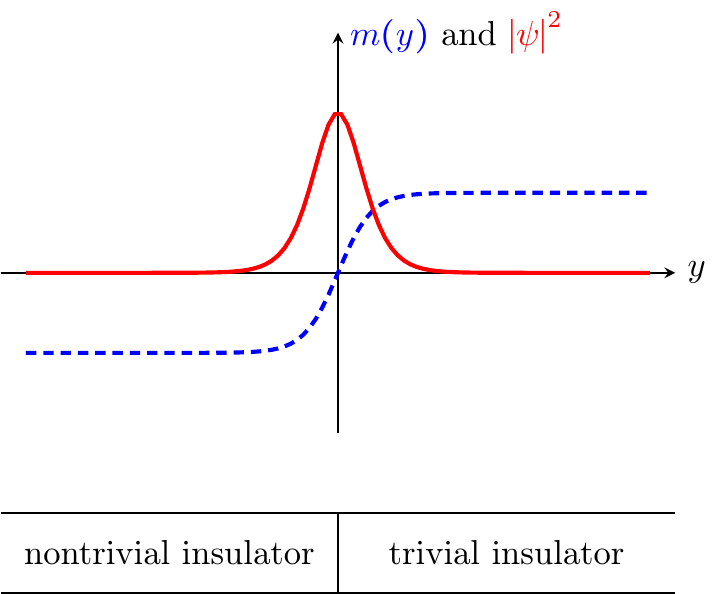}      
\caption{
Schematic view of edge states at a Chern--trivial insulator interface. The mass $m(y)$ (blue dashed line) and the wavefunction amplitude $|\psi|^2$ (red continuous line) are drawn along the coordinate $y$ orthogonal to the interface $y=0$. 
}
\label{fig:ChernEdge}
\end{figure}

 Similarly at the interface between a topological and a trivial insulator for the Kane-Mele model (\ref{eq:KM-Hamiltonian},\ref{eq:defGamma}) 
 the function $d_1(k)$ changes sign at one of the $\lambda_i$ TRIM points. Let us choose 
$m (y)=d_{1}[k=\lambda_{0}](y)$ with $m (y>0)>0$ and $m (y<0)<0$. 
 The functions $d_{i\geq 2}$  are odd around $\lambda_{0}$ and up to a rotation of the local coordinates on the Brillouin zone we have 
 $d_{3}(q)=q_{x}$ and $d_{2}(q)=-q_{y}$ and 
 the linearized Hamiltonian around the TRIM $\lambda_{0}$ reads, 
\begin{equation}
H_{\text{l}}(q) = q_x ~ \Gamma_{3} - q_{y} ~\Gamma_{2} + m (y) ~ \Gamma_{1} .
\end{equation}
This Hamiltonian can be block-diagonalized in the basis (\ref{eq:TensorProductBasis}) leading to the two surface solutions 
\begin{subequations}
\begin{equation}
\psi_{q_{x},\uparrow}(x,y) \propto \ee^{- \ii q_{x} \, x} \, \exp \left[ - \int_{0}^{y} m(y') \, \dd{}y' \right] \; 
\begin{pmatrix}
  0 \\
  1 \\
  0 \\
  0
\end{pmatrix}
\end{equation}
\begin{equation}
\psi_{q_{x},\downarrow}(x,y) \propto \ee^{+ \ii q_{x} \, x} \, \exp \left[ - \int_{0}^{y} m(y') \, \dd{}y' \right] \; 
\begin{pmatrix}
  0 \\
  0 \\
  0 \\
  1
\end{pmatrix}, 
\end{equation}
\end{subequations}
which consist of a Kramers pair of counter propagating states with opposite spins. 
\begin{figure}[ht]
\centering  
\includegraphics[width=8cm]{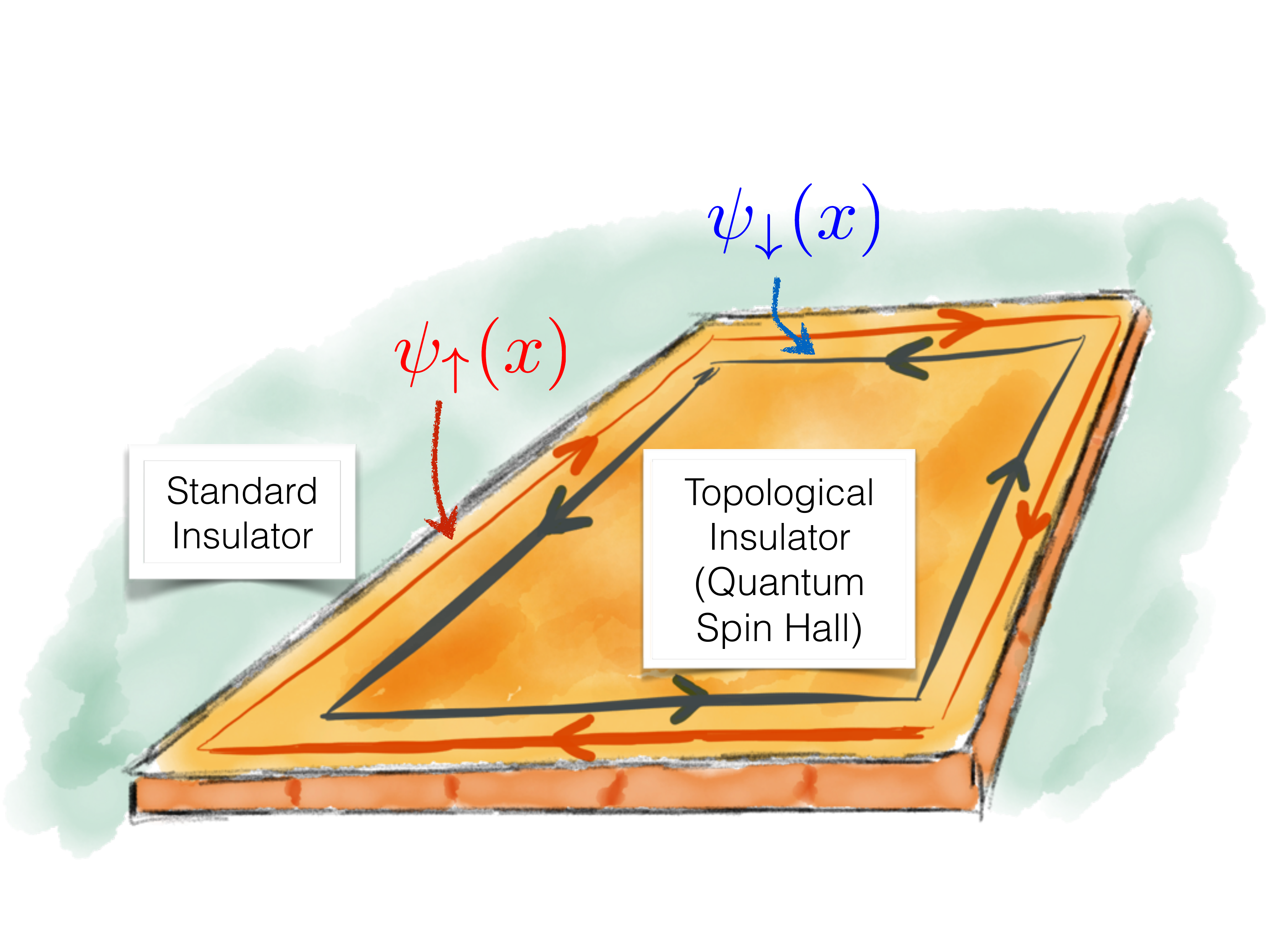}      
\caption{
Illustration of edge states at the interface between a two dimensional Kane--Mele insulator (Quantum Spin Hall Phase) and a standard insulator. 
 A Kramers of counter-propagating states are present, represented here for simplicity by spin-up states (in red) and spin-down states (blue) .
}
\label{fig:Z2Edge}
\end{figure}
A schematic representation of such a pair of edge states is represented in Fig.~\ref{fig:Z2Edge}. 
A mathematical discussion on this bulk-edge correspondence that goes far beyond the present introduction can be found in 
\cite{GrafPorta2012}.

\subsection{From Topology in the Bulk to Critical Semi-metals}

The previous discussion also applies to the transition between bulk phases as a function of an external parameter such as chemical doping, 
instead the distance to an interface. 
We have already encountered this situation in the analysis of the Haldane model (\ref{eq:HaldaneHamiltonianFirstQuantized}) : on the phase diagram of 
 Fig.~\ref{fig:HaldanePhaseDiagram} the critical lines correspond to a situation where the gap $h_z$ 
 closes at one point of the Brillouin zone, and thus correspond to two dimensional Dirac semi-metallic phases. 
 One special point possesses two Dirac points : the case of graphene $M=\phi=0$ where both parity and time-reversal symmetry are restored. 
 A similar scenario holds at the transition from a topological insulator to an ordinary insulator both in two and three dimensions : as such a transition, the 
 coefficient $d_1(\lambda_i)$ of the Kane-Mele model (\ref{eq:KM-Hamiltonian}) vanishes. At the transition, the effective hamiltonian describing the 
 low electronic degree of freedom of the model is obtained by  
 linearizing all three functions $d_i(\lambda_i + q)$ to first order in $q$. Up to dilatations and a rotation in momentum space the Hamiltonian reads : 
 \begin{equation}
 H_\text{Dirac} = \Gamma_1 ~ q_x + \Gamma_2  ~ q_y + \Gamma_3  ~ q_z  , 
 \label{eq:Dirac3D}
 \end{equation}
where the $\Gamma_i$ matrices satisfy a Clifford algebra : $\{ \Gamma_{i} , \Gamma_{i}\} = 2 \delta_{i,j}$ : this is a massless three dimensional Dirac 
equation. 
Hence the critical model at the transition between a topological and an ordinary insulator is  a model of three dimensional massless Dirac particles. 
 This result is not longer valid when inversion symmetry is broken : the transition occurs then in two steps with a critical phase in 
 between~\cite{Murakami:2007}. This critical region corresponds to a semi-metal where only two bands cross at two different points $K,K'$. Around 
 each of these points, an effective $2\times 2$ hamiltonian is sufficient to describre the low energy electronic states, parametrized by Pauli matrices : 
 \begin{equation}
 H_\text{Weyl} = \sigma_1 ~ q_x + \sigma_2  ~ q_y + \sigma_3  ~ q_z . 
  \label{eq:Weyl}
 \end{equation}
 Such a hamiltonian does not describe anymore excitations satisfying a Dirac equation : it is instead called a Weyl hamiltonian. 
Weyl hamiltonians, which have been proposed in particular as low energy hamiltonian in correlated materials \cite{Turner:2013}. They possess remarkable 
surface states with unusual Fermi arcs. 

Much of the recent activity in this direction have focused on the possibility to stabilize a three dimensional 
Dirac semi-metal by crystalline symmetries to realize a three dimensional analog of graphene. 
Indeed, a Dirac point described by a Hamiltonian (\ref{eq:Dirac3D}) can be viewed as the superposition of two Weyl points such as eq.~(\ref{eq:Weyl}) with 
opposite Chern numbers (see below). To prevent the annihilation of the two Weyl points, additional symmetries are required to 
 forbid any hybridization between the associated bands (see \cite{Young:2012} for a general discussion). Such a Dirac phase was recently proposed 
 in Cd$_3$As$_2$ and tested experimentally   \cite{Wang:2013}.

\subsection{Topological Properties of Semi-metallic phases }

\begin{figure}[ht]
\centering  
\includegraphics[width=6cm]{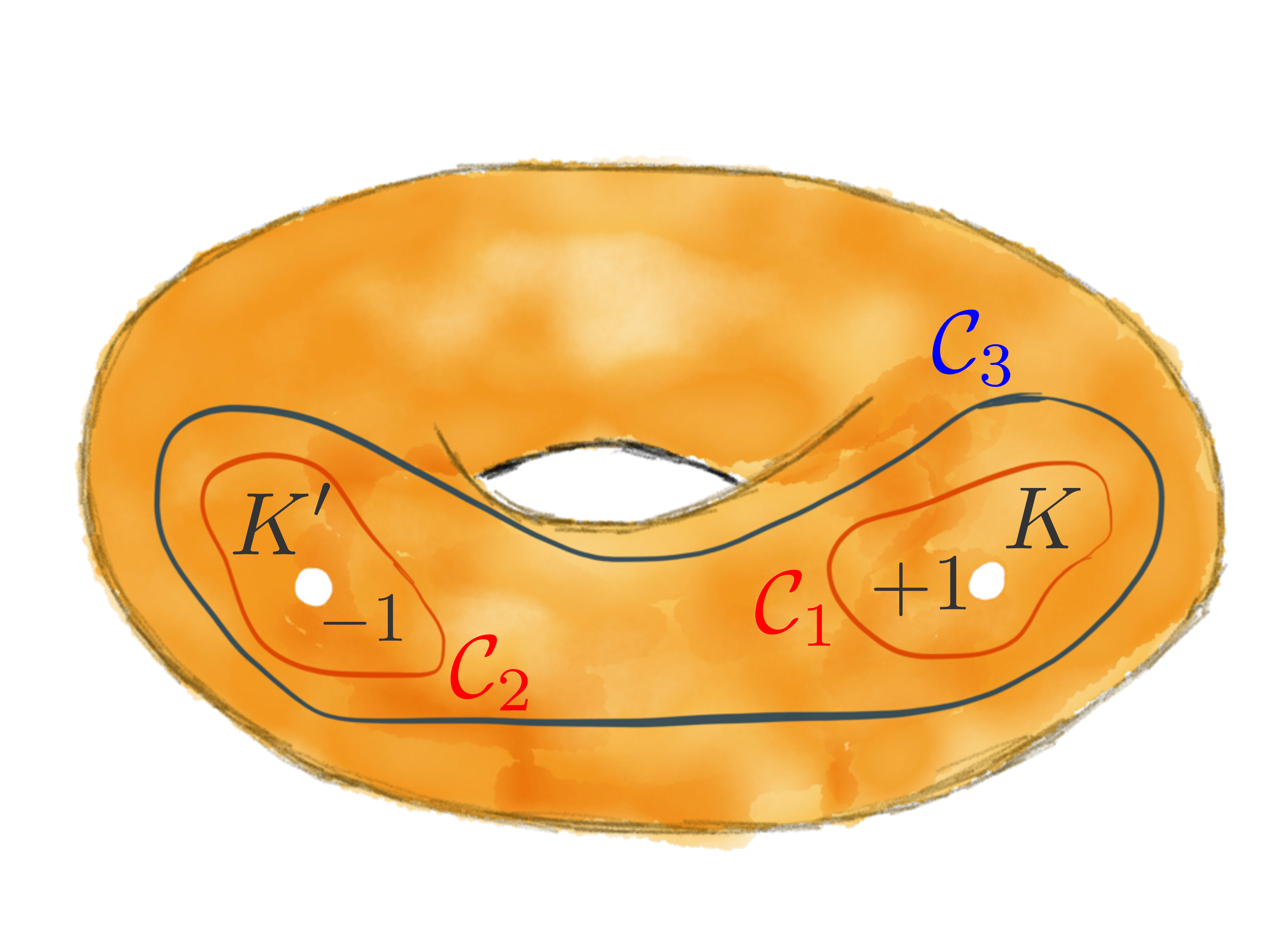}      
\caption{
When the locations of the band crossing  points are excluded from the Brillouin torus,  
new classes of non contractible loops (in d=2) exists, here represented by $\mathcal{C_1}, \mathcal{C_2}$. They allow to define topological charges 
associated with these crossing points. The total charge over all crossing points necessarily vanishes : a loop such as 
$\mathcal{C_3}$ encircling all points is necessary contractible and can deformed to a point. 
}
\label{fig:DiracTopology}
\end{figure}

 Much of the recent theoretical discussions of the stability and existence of semi-metallic phases have relied on a topological characterization of these phases, 
 extenuating the previously discussed characterization of insulators. 
Indeed, we have already encountered that crossing between two bands appeared as analogs of Berry monopoles : 
in the analysis of the Chern index in section \ref{sec:TwoBandModel},  the point $h=0$ corresponding to the two bands crossing
acts as the source of the flux integrated on the surface spanned by $h(k)$, and whose total flux constitute the Chern number. 
 Similarly, the expression (\ref{eq:BerryCurvature2}) of Berry curvature is found to be singular when two bands approach each other. 
 Indeed, it is possible to attribute a topological number to such a band crossing points. 
This classification of topological properties of semi-metals proceeds along the same lines as the classification of defects in orderered media  
(vortices, dislocations, skyrmions, etc.) \cite{Toulouse:1976} : see in particular \cite{Volovik:2003} for a related discussion. 
By definition in such a semi-metal a discrete set of points separate valence from conduction bands. 
If we exclude these points from the Brillouin zone, we obtain two well-separated sub-bundles of valence and conduction bands states, 
similarly to the insulating case. These vector bundle are now defined on the Brillouin zone punctured by these points. 
These two sub-bundles possess non-trivial topological properties due to the presence of these points : due to the modified topology of the punctured Brillouin zone, 
there now exist non-contractible loops ($d=2$) or surfaces ($d=3$) which encircle these points (see Fig~\ref{fig:DiracTopology}). 
 While the Berry curvature of the valence band might vanish everywhere in the punctured Brillouin zone, the Berry parallel transport of eigenstates 
 along a path can be non trivial, associated with a winding independent of the path around the point. 

 In two dimensions,  we can define a winding number associated with any loop $\mathcal{C}$ encircling once the crossing point K, defined by 
\begin{equation}
w_{K} = \frac{1}{\pi} \oint_\mathcal{C} A, 
\end{equation}
where $A$ is the Berry connexion form of the valence band. 
Note that necessarily, the sum of winding numbers of the different crossing points in the Brillouin zone must vanish : it corresponds to the 
  winding number along a path encircling all points, which is necessarily contractible (see Fig~\ref{fig:DiracTopology}).
   In the case of graphene, the two Dirac points are associated with opposite winding numbers $\pm 1$. These numbers are topological in the following sense : they are robust with respect to any perturbation that doesn't open a gap at a Dirac point. From the Haldane model (\ref{eq:HaldaneHamiltonianFirstQuantized}) we can identify these perturbations as those preserving time-reversal and parity symmetry. Within the ensemble of such perturbations, the topological numbers associated with each point is robust. The only way to modify the topology of the band structure is through the merging or fusion of these points. Such topological transitions have been studied in the context of graphene in \cite{Montambaux:2009} and observed in cold atoms lattices \cite{Tarruell:2012}. 
 Similar reasoning allows to define the topological number characterizing a semi-metallic point $K$  in three dimensions as the Chern number 
\begin{equation}
n_{K} = \frac{1}{2\pi} \oint_\mathcal{S} F  
\end{equation}
where $\mathcal{S}$ is now a surface enclosing the point $K$, and the Berry curvature is defined on the valence bands. 
 For a model with two bands crossing, described by a general Weyl Hamiltonian 
 $H_\text{Weyl} (k=K+q) =  v_{ij} q_i \sigma_j$, the calculation goes exactly along the lines of the analysis of  section \ref{sec:TwoBandModel}, 
 except that the integral over momentum $q$  does not run over but around the two dimensional sphere encircling $K$ instead of the 
 two dimensional Brillouin torus. We obtain a Chern number given by  \cite{Young:2012} 
 $n(K) = \textrm{sign} \left( \det \left[ v_{ij} \right] \right) $.

\section{Conclusion and Perspectives}

At the point of concluding these notes, it is worth mentioning a few subjects that have not been covered here by lack of time or knowledge. 
The ideas sketched in these notes have led to a classification of all $10$ topological properties of gapped phases, whether insulating or superconducting, 
within a single particle framework. Such a classification can be obtained either by studying the robustness of surface states with respect to disorder, or 
by studying directly the topological classes of vector bundles within K-theory.  Of great natural interest is the incorporation of interactions within this 
picture. Moreover, in an attempt to describe the origin of  topological properties of bands, we have neglected the description of the essentials  properties 
of the associated Dirac surface states and thus the physical properties of these materials. This constitute a vast subject, and the interested reader should 
refer to the serious reviews already mentioned  \cite{HasanKane2010, Qi:2011,Bernevig,FranzMolenkamp}. 

\medskip
\noindent{\bf Acknowledgment :}  I thank Pierre Delplace, Michel Fruchart, Thibaut Louvet and above all Krzysztof Gawedzki,  from whom I've learned  through
 friendly and stimulating discussions most of what is included in this review. 

\appendix 

\section{Two useful trivializations of the Bloch Bundle}
\label{sec:appendix}

\subsection{Trivialization by Fourier Transform}
An example of  trivialization of the Bloch bundle $\mathcal{H}$ is provided by starting from a choice of unit cell 
$\mathcal F $ of the crystal $\mathcal{C}$ (see section \ref{sec:Bloch}). Such a unit cell is constituted of $N$ points 
 $x_\alpha, \alpha=1,\dots,N$.
  Let us denote  by $|e^{\un }_\alpha (k) \rangle$ the Fourier transforms of functions $\delta_{x,x_\alpha}$ concentrated at these points $x_\alpha$. 
These Fourier transform constitute a smooth set of sections of the Bloch bundle, which furthermore constitute a basis of each fiber $\mathcal{H}_k$ : 
 Bloch states decompose uniquely as $| \varphi (k)\rangle=\sum_\alpha \varphi(k;x_\alpha) |e^{\un }_\alpha (k)\rangle $. 
 We shall denote by $\nabla^{\un}$ the flat connection associated to the trivialization $k\mapsto e^{\un i}_k$ of $\mathcal{H}$. 
 This trivialization of $\mathcal{H}$ and its associated connection depends on the associated choice of unit cell $\mathcal F$.
Another choice $\mathcal F'$ of unit cell is related to  $\mathcal F$  up to a possible relabeling of points by 
$x'_\alpha=x_\alpha+\gamma_\alpha$ with $\gamma_\alpha$ a vector of the Bravais lattice $\gamma_\alpha\in\Gamma$. 
 Both trivialization are then related by the simple proportionality relation $| e'^{\un}_\alpha (k) \rangle=\ee^{\ii k\cdot\gamma_\alpha} | e^{\un }_\alpha (k)\rangle $, 
 and the associated flat connexions only differ (as expected for two flat connexions) by a  differential 1-form : 
\begin{equation}
{\nabla'^{\un}}=\nabla^{\un}-\ii\sum_\alpha   |e^{\un}_\alpha (k) \rangle\langle e^{\un}_\alpha (k)| \,dk\cdot\gamma_\alpha  . 
\end{equation}

\subsection{Trivialization from Periodic Functions}
 A second choice of trivialization appears more canonical as it depends in a trivial way of the arbitrariness of this definition (the choice of origin of space). 
 It corresponds to the common writing of Bloch functions $|\varphi (k) \rangle \in \mathcal{H}_k$ as a  functions of periodic functions 
\begin{equation}
\varphi (k;x)=\ee^{-\ii k\cdot(x-x_0)}u (k;x) , 
\label{eq:PF}
\end{equation}
 where $x$ is any point in the crystal $\mathcal{C}$ and $x_0$ an arbitrary origin of the  Euclidean space, not necessarily in the crystal. 
In this expression,  $u(k;.)$ is a periodic function on the Bravais lattice : $u(k;x+\gamma)=u(k;x)$ for $\gamma\in\Gamma$. 
 Hence eq.~(\ref{eq:PF}) establishes a relation between each fiber $\mathcal{H}_k$  of quasi-periodic Bloch functions and the vector space $\ell^2(\mathcal C/\Gamma)$ 
 of $\Gamma$-periodic function on the crystal $\mathcal{C}$. 
Such a function $\varphi (k;x)$  does not possess the periodicity of the reciprocal lattice as   
$u (k+G;x) =\ee^{\ii G\cdot(x-x_0)}u(k;x)$ for  $G\in\Gamma^\star$.
Hence the writing (\ref{eq:PF}) establishes an identification between the Bloch bundle $\mathcal{H}$
and the quotient of the trivial bundle $\mathbb{R}^d\times \ell^2(\mathcal  C/\Gamma)$ of periodic function $u$ indexed by a real vector 
$k\in \mathbb{R}^d$, by the action of the reciprocal lattice $\Gamma^\star$ (which preserves the scalar produce on $\ell^2(\mathcal C/\Gamma)$) : 
\begin{equation}
(k,u(x))\longmapsto(k+G,\,\ee^{\ii G\cdot(x-x_0)}u(x)) , \textrm{ for } G \in \Gamma^\star . 
\label{eq:ActionGamma*}
\end{equation}
The bundle $\mathbb{R}^d\times \ell^2(\mathcal C/\Gamma)$ of periodic function $u$ indexed by a real vector  
is trivial in the sense defined previously : it possesses a natural flat connection obtained by choosing a basis of periodic functions  
$\ell^2(\mathcal C/\Gamma)$ independent of $k$. 
A natural choice for this basis is 
 $u^\alpha(x)=\sum_{\gamma\in\Gamma} \delta_{x,x_\alpha+\gamma}$, $\alpha=1,\cdots,N$ where the $x_\alpha$ are the points of a unit cell 
$\mathcal F$. It should be noted that the $u^\alpha$ are independent of this choice of $\mathcal F$ : they are indexed by sub-lattices, and not by a 
choice of point $x_\alpha$ in each sub-lattice. 
 This basis is pulled back as a basis of Bloch functions in each $\mathcal{H}_k$ as 
$e^\deux_\alpha(k;x)=\ee^{-\ii k\cdot(x-x_0)}u^i(x)$
 The connexion $\nabla^\deux$ is now defined by the exterior derivative of its sections in this basis (see eq.(\ref{eq:FlatConnexion})).


\end{document}